\documentclass[authoryear, review]{elsarticle}
\usepackage[utf8]{inputenc} 
\usepackage{geometry}
\usepackage{float}
\usepackage[hidelinks]{hyperref}
\usepackage{xcolor}
\usepackage[bitstream-charter]{mathdesign}

\usepackage{amssymb}
\usepackage{amsmath} 
\usepackage{mathtools}
\usepackage{array}
\usepackage{bbm} 
\usepackage{breqn} 
\usepackage{booktabs}
\usepackage{epstopdf, epsfig}
\usepackage{subcaption}
\usepackage{appendix}
\usepackage{natbib}

\makeatletter
\def\ps@pprintTitle{%
 \let\@oddhead\@empty
 \let\@evenhead\@empty
 \def\@oddfoot{\centerline{\thepage}}%
 \let\@evenfoot\@oddfoot}
\makeatother

\begin{document}

\begin{frontmatter}
\title{Capital Demand Driven Business Cycles: Mechanism and Effects} 
\author[1,2,3]{Karl Naumann-Woleske\corref{cor1}}
\ead{karl.naumann@polytechnique.edu}
\author[1,2,4]{Michael Benzaquen}
\ead{michael.benzaquen@polytechnique.edu}
\author[5]{Maxim Gusev}
\ead{maxim.gusev@lgtcp.com}
\author[5]{Dimitri Kroujiline}
\ead{dimitri.kroujiline@lgtcp.com}
\cortext[cor1]{Corresponding author}
\address[1]{Chair of Econophysics and Complex Systems, Ecole Polytechnique, 91128 Palaiseau Cedex, France}
\address[2]{LadHyX, UMR CNRS 7646, Ecole Polytechnique, 91128 Palaiseau Cedex, France}
\address[3]{New Approaches to Economic Challenges Unit (NAEC), OECD, 75016 Paris, France}
\address[4]{Capital Fund Management, 75007 Paris, France}
\address[5]{LGT Capital Partners, 8808 Pf\"affikon, Switzerland}

\begin{abstract}

We develop a tractable macroeconomic model that captures dynamic behaviors across multiple timescales, including business cycles. The model is anchored in a dynamic capital demand framework reflecting an interactions-based process whereby firms determine capital needs and make investment decisions on a micro level. We derive equations for aggregate demand from this micro setting and embed them in the Solow growth economy. As a result, we obtain a closed-form dynamical system with which we study economic fluctuations and their impact on long-term growth. For realistic parameters, the model has two attracting equilibria: one at which the economy contracts and one at which it expands. This bi-stable configuration gives rise to quasiperiodic fluctuations, characterized by the economy’s prolonged entrapment in either a contraction or expansion mode punctuated by rapid alternations between them. We identify the underlying endogenous mechanism as a coherence resonance phenomenon. In addition, the model admits a stochastic limit cycle likewise capable of generating quasiperiodic fluctuations; however, we show that these fluctuations cannot be realized as they induce unrealistic growth dynamics. We further find that while the fluctuations powered by coherence resonance can cause substantial excursions from the equilibrium growth path, such deviations vanish in the long run as supply and demand converge.

\end{abstract}

\begin{keyword}
Business Cycles \sep Economic Growth \sep Interactions-Based Models \sep Dynamical Systems \sep Limit Cycle \sep Coherence Resonance 
\end{keyword}
\end{frontmatter}
\newpage
\section{Introduction}

\begin{quotation}
	My view was, and still is, that the most urgent current analytical need was for a way of fitting together short-run macroeconomics, when the main action consists of variations in aggregate demand, with the long run factors represented by the neoclassical growth model, when the main action is on the supply side. Another way of saying this is that short-run and long-run models of macroeconomic behavior need a way to merge in a practical macroeconomics of the medium run \citep{Solow2005}.
\end{quotation}

The field of economics has long been aware of a conceptual dichotomy between studies of short-term dynamics and models of long-term growth. An early distinction was made between the Hicks IS-LM model \citeyearpar{Hicks1937} and the Solow growth model \citeyearpar{Solow1956}. The developments in both approaches have captured important dynamics at their respective timescales, such as short-term demand effects and endogenous drivers of long-term growth \citep[e.g.][]{AghionHowitt1992}. Yet it is not well understood how the dynamics at different timescales are interlinked and how medium-term disequilibrium dynamics impact the long-term growth trend of the economy.

Since the World War II, the United States of America alone has faced twelve recessions. While the severe short-term consequences of these crises are appreciated, understanding of the long-lasting impact on growth remains underdeveloped. 
The pervasive recurrence of booms and busts has thus sparked research into the linkages between economic volatility and growth \citep{CooleyPrescott1995,AghionHowitt2006, PriesmeierStahler2011, BakasEtAl2019}. Theoretical as well as empirical investigations have turned out to be inconclusive, as authors disagree on both the sign and magnitude of the ultimate effect of volatility on growth.\footnote{
    We suggest \citet{BakasEtAl2019} for a comprehensive review of this literature.
} Theoretical literature is divided into two dominant strands that stem from either Schumpeterian notions, in which volatility is good for growth \citep[based on][]{Schumpeter1939, Schumpeter1942}, or the learning-by-doing concept \citep[based on][]{Arrow1962}, where volatility is detrimental to growth. The conflicting theoretical frameworks and ambiguous empirical findings indicate that new, alternative approaches may be needed to decipher the genuine nature of the relationship between volatility and growth. Current literature does not generally consider the impact of the interactions among economic agents and their collective dynamics on long-term growth. It is this impact and its underlying mechanisms that we seek to capture and explain. 

We are motivated by the micro-to-macro approach of agent-based modeling \citep{LeBaronTesfatsion2008, DawidDelliGatti2018a, HommesLeBaron2018} and, especially, the Keynes-meets-Schumpeter class of models \citep{DosiEtAl2010, DosiEtAl2015} that study the linkages between endogenous growth and demand policy. While agent-based models successfully capture many complex phenomena, they are generally analytically intractable, making the analysis of the precise mechanics linking volatility and growth difficult. Our approach remains distinct as we aim to derive a tractable system of equations for the aggregate dynamics from micro-level interactions.

This paper's objective is to develop a model of capital demand driven economic fluctuations, in which interactions among agents to coordinate on economic outcomes lead to periods of booms and busts, and apply it to examine how fluctuations affect the economy across different timescales and possibly shape its long-term growth. Inspired by \citet{Keynes1936}, our focus on capital demand is motivated by the observation that firms' investment is both pro-cyclical and volatile \citep{StockWatson1999}, suggesting investment decisions play a key role in business cycles. We treat investment decision-making as an interactions-based process whereby firm managers exchange views and affect each other's opinions. In other words, we emphasize strategic complementarity and peer influence that cause managers to coalign their individual expectations at the micro level. We use the framework developed in \citet{GusevEtAl2015} and \citet{KroujilineEtAl2016} to describe this interaction process mathematically and derive the macroscopic equations governing the dynamics of aggregate capital demand. To close the economy while highlighting the demand-driven effects, we attach these equations to a simple supply side component represented by the Solow growth model \citeyearpar{Solow1956}. 

As a result, we obtain a closed-form dynamical system, hereafter the Dynamic Solow model, which enables us to study a broad range of economic behaviors. The model’s primary contribution is the identification of a new mechanism of business cycles that captures their quasiperiodic nature characterized by one or several peaks in a wide distribution of cycle lengths.

We show that, for economically realistic parameters, the Dynamic Solow model admits two attracting equilibria that entrap the economy in either a contraction or expansion.\footnote{The 2008 crisis gave new impetus to revisiting the single equilibrium framework; e.g. \cite{VinesWills2020}  recently made the case for moving towards a multi-equilibrium paradigm.
} The equilibria are indeterminate \citep{BenhabibFarmer1999} as both the path to and the choice of equilibrium depend on the beliefs of the agents themselves. The entrapment is asymmetric because technological progress, introduced externally, causes the economy to stay on average longer in expansion than contraction, contributing to long-term growth. The flow of exogenous news continually perturbs the economy stochastically and prevents it from settling at either equilibrium. Over time, the economy tends to drift slowly towards the boundary between the contraction and expansion regions, making it easier for a news shock to instigate a regime transition in line with the “small shock, large business cycle” effect \citep{BernankeEtAl1996}. This endogenous mechanism generates quasiperiodic fluctuations as it involves both deterministic dynamics and stochastic forcing.

Such a mechanism, whereby noise applied to a dynamical system leads to a quasiperiodic response, is known as coherence resonance \citep{PikovskyKurths1997}. It occurs in situations where the system has long unclosed trajectories such that even small amounts of noise can effectively reconnect them and thus create a quasiperiodic limit cycle. Coherence resonance emerges naturally in bi-stable systems\footnote{In dynamical systems, bistability means the system has two stable equilibrium states.}, including our model.

The coherence resonance mechanism differentiates the Dynamic Solow model from preceding research that has often considered limit cycles as the endogenous source of economic fluctuations.\footnote{
	The early literature comprises \citet{Hicks1937}, \citet{Kaldor1940} and \citet{Goodwin1951}. Later reviews include \citet{BoldrinWoodford1990}, \citet{Scheinkman1990}, \citet{Lorenz1993} and \citet{Gandolfo2009}. We also note \citet{BeaudryEtAl2020} as an influential recent investigation. 
}
In particular, \citet{BeaudryEtAl2020} propose an extended Dynamic Stochastic General Equilibrium model, in which the quasiperiodic character of fluctuations comes from noise acting directly on a periodic limit cycle. We argue, however, that coherence resonance may be the preferred route to generating business cycles as it requires noise only as a catalyst, thus relying much less on random shocks to reproduce regime variability. Furthermore, we show that the fluctuations produced by a noise-perturbed limit cycle, which is as well recovered in a certain parameter range in our model, dampen long-term growth and unrealistically cause capital demand to diverge from supply in the long run.


We note that the Dynamic Solow model nests two limiting cases that match those of previous literature. In the case where capital demand is persistently higher than supply, the model recovers the exponential equilibrium growth of the classic Solow model. In the opposite case, where capital demand is persistently lower than supply, the model exhibits quasiperiodic fluctuations driven by a coherence resonance mechanism similar to that in \citet{KroujilineEtAl2019}.

We explore the Dynamic Solow model numerically across multiple timescales, from months to centuries, and identify business cycles as quasiperiodic fluctuations that most frequently last 40-70 years. These fluctuations may be associated with Kondratieff cycles if interpreted as investment driven.\footnote{Kondratieff himself attributed these cycles to capital investment dynamics. This interpretation was further advanced by a number of papers in the 1980s. Kondratieff cycles are, however, more commonly linked to technological innovation. There have also been attempts to combine investment and innovation explanations. For a review see \citet{KorotayevTsirel2010}.}
\citet{KorotayevTsirel2010} employ spectral analysis to suggest the existence of long-term business cycles. However, the academic community remains divided on this issue and the research has been focused primarily on the fluctuations in the 8-12 year range. These shorter-term cycles cannot emerge in our model because it does not include accelerators such as the financial sector or household debt.

Currently, many macroeconomic models describe an economy in or near equilibrium. Most prominent is the Dynamic Stochastic General Equilibrium class of models \citep[see][for recent reviews]{ChristianoEtAl2018, KaplanViolante2018}. While behavioral limitations and various frictions have been considered, these models operate in an \textit{adiabatic} regime where equilibrium is reached more quickly than the environment changes. In other words, there is some form of perfect coordination (e.g. market clearing where supply and demand equate) among all agents at each point in time. Over long timescales this treatment may be justified, but in the near term coordination failures are inevitable, leading to pronounced fluctuations and persistent spells of disequilibrium. 
 
The Dynamic Solow model enables us to study both the disequilibrium fluctuations and the equilibrium growth. We examine the impact of fluctuations on growth and show that fluctuations can affect economic expansion over extended time intervals. However, the deviations from the balanced growth path disappear with time as demand and supply converge asymptotically in the long run.  

The remainder of this paper is structured as follows. In Section \ref{sec_model} we introduce and explain the mechanics of dynamic capital demand and the Solow growth framework within which it rests. Section \ref{sec_limitcases} considers two limiting cases: first, we obtain the equilibrium growth path when capital demand exceeds supply; and second, we investigate the demand dynamics and highlight the mechanism underlying fluctuations when capital supply exceeds demand. Section \ref{sec_results} formulates and studies the general case of the Dynamic Solow model, focusing on the analysis of mid-term fluctuations and long-term growth. Finally, Section \ref{sec_conclusion} concludes by reflecting on the work done and suggests further avenues of research.

\section{The Dynamic Solow Model}\label{sec_model}
This section develops the Dynamic Solow model.\footnote{M. Gusev and D. Kroujiline formulated and presented the basic ideas behind the model at the Macroeconomic Instability seminars of the "Rebuilding Macroeconomics" project (www.rebuildingmacroeconomics.ac.uk) run by the National Institute of Economic and Social Research in London \citep{GusevKroujiline2020a}.} The modeling framework is set out in Section \ref{sec_model_structure} and the equations of the model components are derived in Sections \ref{sec_model_production}-\ref{sec_model_capmkt}.\footnote{The code for this model is written in \texttt{Python} and is available at \href{https://github.com/KarlNaumann/DynamicSolowModel}{github.com/KarlNaumann/DynamicSolowModel}.}

\subsection{Model Structure}\label{sec_model_structure}
\begin{figure}[htb!]
    \centering
    \includegraphics[width=0.9\textwidth]{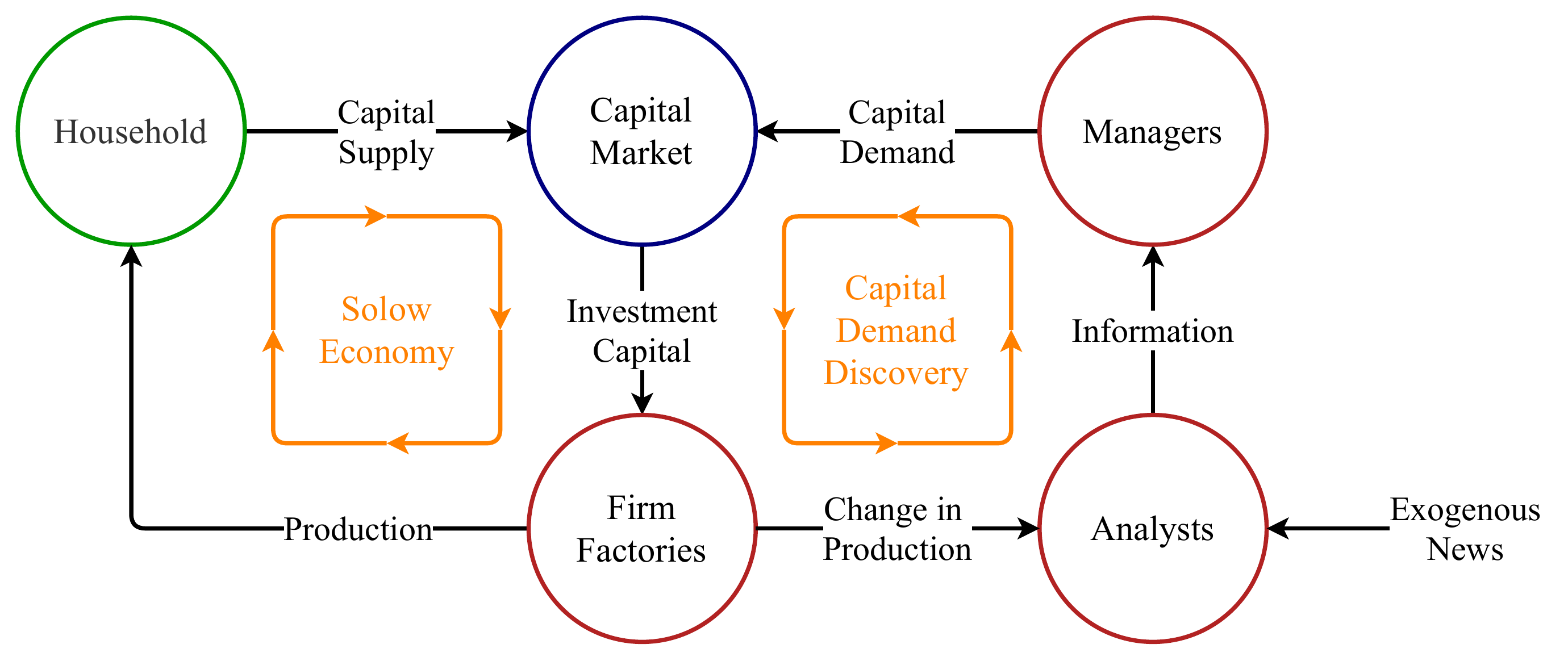}
    \caption{A conceptual flowchart of the Dynamic Solow model. Each individual circle depicts an entity or agent. The circles' color indicates whether they belong to the same entity (notably, firm  managers and firm factories are both parts of the firm). Labeled black arrows define the flows between the respective entities or agents. The orange loops highlight (i) the Solow economy (left loop) and (ii) the dynamic demand decision-making process introduced in this section (right loop).}
    \label{fig_solow_flowchart}
\end{figure}

The Dynamic Solow model is illustrated in Figure \ref{fig_solow_flowchart}. It consists of a dynamic demand framework that we propose to describe how firms determine capital needs and make investment decisions (right loop), to which we attach the familiar circular income flow of the Solow growth economy (left loop).\footnote{
	We choose this supply-side framework for the following reasons: (i) the capital supply dynamics are less important on the timescales where we expect to find fluctuations and thus can be modeled approximately; (ii) the assumption that households save a constant fraction of income is an appropriate leading-order approximation since it is the first term in the Taylor series expansion of savings as a general function of income; and (iii) the Solow model is a parsimonious representation of economic growth, sharing the basics with many macroeconomic models, which may be helpful to extending our approach to more sophisticated settings.
} 

In the Solow economy, households supply capital through savings and firms convert capital into output for household consumption. Both households and firms are static decision-makers. Households save a fixed share of income and firms convert all supplied capital into production. In contrast, we aim to describe how firms develop a strategic business outlook based on their reading of the current economic situation and accordingly determine their capital needs so as to adjust production capacity. Firms thus become active decision-makers, which results in a dynamically evolving capital demand.

Organizational decision-making is a complex process with competing goals and targets, often based on industry-standard business planning and operating procedures \citep{CyertMarch1992, MillerCardinal1994}. Without needing to make firm goals explicit, we posit that corporate decision-making can be viewed as a composite of two distinct processes occurring on different timescales. First, there is information gathering and analysis, characterized by the frequency with which exogenous information such as ad-hoc company news, monthly statistics releases or quarterly earnings reports becomes available. Second, there is the formation of firms’ expectations about the future based on the analysis of collected information, which is then translated into investment decisions. Note that the strategic aspect of investment decision-making implies longer timescales than those of information gathering and analysis.

We model this two-tier decision-making on the microscale by introducing two classes of agents: analysts who collect and analyze relevant information and managers who use this analysis to develop a business outlook and make investment decisions. There are industries where these two classes of agents actually exist (e.g. analysts and investors in finance), whereas in other situations this division serves as a metaphor for the different actions completed by an individual participant. Our objective is to derive the macro-level equations for aggregate demand from this micro setting. 

External information enters the decision-making process at the analyst level. Initially, we may neglect the cost side and focus solely on revenue generation, elevating in relevance the expectation of future consumption. Motivated by recent work on extrapolative beliefs in finance \citep{GreenwoodShleifer2014, KuchlerZafar2019, DaEtAl2021}, we assume that analysts base their expectations primarily on the current state of the economy by extrapolating the consumption growth into the future. As such, we wish to carve out consumption growth as the most relevant information stream and model all other news as exogenous noise \citep[treating news shocks similarly to][]{AngeletosLaO2013,AngeletosEtAl2020,BeaudryPortier2014}. We can further simplify by replacing consumption with production since consumption is approximated as a constant fraction of production in the model. The resulting system acquires a feedback mechanism as higher output growth leads to increasing expectations that cause greater investment, inducing further increases in output growth and starting the process anew.

On the manager level, we emphasize the impact of the opinions and actions of competitors on decision-making, following the growing body of research on peer influence in business \citep{GriskeviciusEtAl2008} and strategic complementarity \citep{CooperJohn1988, BeaudryEtAl2020}. More specifically, we assume that managers exchange views within their peer network with the purpose of coaligning their expectations about the economy. 

The Dynamic Solow model employs, as discussed, two different processes for capital demand and supply: firms determine capital needs dynamically via individual interactions and economic feedback while households supply capital in proportion to income. Thus, demand and supply do not necessarily match at each point in time, which brings us to the discussion of capital market clearing on different timescales. The dynamic demand discovery process occurs on timescales much shorter than the timescale of technological growth. At these short and intermediate timescales -- relevant to information gathering, investment decision-making and production adjustment -- prices are rigid and we expect demand and supply to behave inelastically. However, over long time horizons in which the economy is advancing along the equilibrium growth path, prices become flexible and the capital market clears via price adjustment. Therefore, we expect that demand and supply converge in the long run.

As such, the conceptual framework behind the model is now complete. The remainder of Section \ref{sec_model} is as follows. Section \ref{sec_model_production} extends the usual equation for aggregate economic production to include the (shorter) timescales at which production capacity adjusts. Section \ref{sec_model_household} briefly introduces a representative household and the capital motion equation. Section \ref{sec_model_demand} derives the equations for aggregate capital demand from the micro-level agent-based formulation outlined above. Finally, Section \ref{sec_model_capmkt} sets out conditions for capital market clearing.

\subsection{Production}\label{sec_model_production}
We represent aggregate output by a Cobb-Douglas production function that takes invested capital as an input,\footnote{
	For simplicity, we normalize the initial technology level to unity and do not consider labor, which is equivalent to normalizing the labor component to unity or taking the variables in per capita terms.
} generically written as 
\begin{equation}\label{eq_cd_basic}
    Y = e^{\varepsilon t}K^\rho,
\end{equation} 
with output $Y$, invested capital $K$, capital share in production $\rho$ and technology growth rate $\varepsilon$. 
Equation \eqref{eq_cd_basic} implies that output adjusts immediately to any change in capital. In other words, it is only valid on timescales longer than the time it takes to adjust the production capacity (e.g. the construction of a new factory or installation of new machinery). Since we are also concerned with decision-making processes that occur at much shorter timescales than production adjustment, we introduce a dynamic form of production 
\begin{equation}\label{eq_y_dot_first}
    \tau_y\dot{Y} = -Y + e^{\varepsilon t} K^\rho ,
\end{equation}
where the dot denotes the derivative with respect to time and $1\ll\tau_y\ll1/\varepsilon$ is the characteristic timescale of production capacity adjustment.\footnote{
	In reality, the adjustment periods are asymmetric because it is easier to reduce capacity than to increase it. For simplicity, we treat capacity increases and decreases as if they were symmetric.
} In the short run, this equation describes the dynamic adjustment of output to new capital levels. In the long run, we recover the Cobb-Douglas production form \eqref{eq_cd_basic} as $\tau_y\dot{Y}$ becomes negligibly small for $t\gg\tau_y$.

Finally, we rewrite equation \eqref{eq_y_dot_first} with log variables $k=\ln K$ and $y=\ln Y$ as
\begin{equation}\label{eq_y_dot}
    \tau_y \dot{y} = e^{\rho k + \varepsilon t - y} - 1.
\end{equation}

\subsection{Households and Capital Supply}\label{sec_model_household}
We consider a single representative household that is the owner of the firm and thus receives $Y$ as income. A fixed proportion of income, expressed as $\lambda Y$, is saved and the remainder is consumed. This is a convenient simplification that allows us to focus on the effects of dynamic capital demand. A constant savings rate can also be viewed as a leading-order Taylor expansion of household savings as a general function of income, making it a sensible first approximation.

The total savings are available to firms to invest. We denote them as capital supply $K_s$. The working capital used in production, $K$, suffers depreciation at a rate $\delta$. As households are the owners of the capital, the loss $\delta K$ is attributed to the capital supply. Consequently, the supply dynamics take the form 
\begin{equation}\label{eq_cap_sup}
    \dot{K}_s = \lambda Y - \delta K.
\end{equation}

Setting $k_s=\ln K_s$, we reformulate equation \eqref{eq_cap_sup} using log variables as
\begin{equation}\label{eq_ks_dot}
    \dot{k}_s = \lambda e^{y-k_s} - \delta e^{k-k_s}.
\end{equation} 

\subsection{Dynamic Capital Demand}\label{sec_model_demand}
In this section, we derive the equations for aggregate capital demand. As set out in Section \ref{sec_model_structure}, this derivation is based on a micro-level framework that divides the firms’ investment planning into two processes occurring at different speeds: fast-paced information gathering and analysis; and slow-paced decision-making. We model these processes 
with two classes of interacting agents: analysts who collect and analyze relevant information; and managers who use this analysis to develop their strategic business outlook and make investment decisions.\footnote{This modeling approach adapts the investor-analyst interaction framework, developed for the stock market in \citet{GusevEtAl2015} and \citet{KroujilineEtAl2016}, to the macroeconomic context.}

In mathematical terms, we consider two large groups of agents: analysts $i\in\{1,\dots,N_{h}\}$ and managers $j\in\{1,\ldots,N_{s}\}$, where $N_{h}\gg1$ and $N_{s}\gg1$. Each analyst and manager has a positive or negative expectation about the future path of production, respectively $h_{i}=\pm1$  and  $s_{j}=\pm1$. The agents interact by exchanging opinions. As a result, the agents influence each other's expectations and tend to coalign them. To stay general, we assume analysts and managers interact among themselves and with each other. These individual interactions drive the evolution of the macroscopic variables: average analyst expectation $h$ (information) and average manager expectation $s$ (sentiment). 

At each moment of time $t$, sentiment $s$ is given by
\begin{equation}\label{eq_s_first}
    s(t) =n_{+}(t) -n_{-}(t),
\end{equation}
where $n_{+}={N_{s}^{+}}/{N_{s}}$ and $n_{-}={N_{s}^{-}}/{N_{s}}$, with $N_{s}^{+}$ and $N_{s}^{-}$ representing the respective number of optimists $(s_{j}=1)$ and pessimists ($s_{j}=-1$). By construction, $s$ varies between $-1$ and $1$. At the leading order, we treat interaction as though each $s_j$ is affected by the collective opinions $s$ and $h$ (similarly constructed), each forcing $s_j$ in their respective directions.\footnote{
	This treatment, known as the mean-field approach, is the leading-order approximation for a general interaction topology.
} As a result of this simplification, we can introduce the total force of peer influence $F_s$ acting on each manager as 
\begin{equation}\label{eq_demand_force_s}
F_{s}(s,h) = \beta _{1}s(t) + \beta _{2}h(t) +E_{s}(t),
\end{equation}
where $\beta _{1}>0$ and $\beta _{2}>0$ are the sensitivities and $E_s$ denotes general exogenous influences (to be specified later). Equation \eqref{eq_demand_force_s} implies that as the collective expectations of managers and analysts grow more optimistic, the stronger the force exerted on a pessimistic manager to reverse her views (and vice versa).

In addition, managers may be affected by a multitude of idiosyncratic factors causing them to occasionally change opinions irrespective of other participants. We treat them as random disturbances and, accordingly, introduce the transition rates $p^{-+}$ as the probability per unit time for a manager to switch from a negative to positive opinion and $p^{+-}$ as the probability per unit time of the opposite change. We can express the changes in $n_+$ and $n_-$ over a time interval $\Delta t$ as
\begin{eqnarray}\label{eq_n_transition}
n_+(t+\Delta t) &=& n_+(t) + \Delta t\left(n_-(t)p^{-+}(t) - n_+(t)p^{+-}(t)\right)\label{eq_n_plus},\\
n_-(t+\Delta t) &=& n_-(t) + \Delta t\left(n_+(t)p^{+-}(t) - n_-(t)p^{-+}(t)\right)\label{eq_n_neg}.
\end{eqnarray}
Noting that $n_+=(1+s)/2$ and $n_-=(1-s)/2$, we subtract \eqref{eq_n_neg} from \eqref{eq_n_plus} to obtain in the limit $\Delta t \rightarrow 0$ 
\begin{equation}\label{eq_s_rates}
\dot{s} = (1-s)p^{-+} - (1+s)p^{+-}.
\end{equation}

To complete the derivation, we must find out how the transition rates depend on peer influence: $p^{-+}=p^{-+}(F_s)$ and $p^{+-}=p^{+-}(F_s)$. It follows from \eqref{eq_n_transition} that in the state of equilibrium, when $n_\pm(t+\Delta t)=n_\pm(t)$, the condition ${p^{-+}}/{p^{+-}}={n_+}/{n_-}={N_{s}^+}/{N_{s}^-}$ holds. Thus ${p^{-+}}/{p^{+-}}$ can be interpreted as the ratio of optimists to pessimists. We can assume this ratio changes proportionally to a change in $F_s$, that is $d{\left(N_{s}^+/N_{s}^-\right)}/\left(N_{s}^+/N_{s}^-\right)=\alpha d F_s$ where $\alpha$ is a positive constant. This interpretation allows us to write $d{\left(p^{-+}/p^{+-}\right)}/\left(p^{-+}/p^{+-}\right)=\alpha d F_s$, which leads to
\begin{equation}\label{eq_force}
	\frac{p^{-+}}{p^{+-}}=e^{\alpha F_s}.
\end{equation}
Condition \eqref{eq_force} implies correctly that $p^{-+}>p^{+-}$ for $F_s>0$, $p^{-+}=p^{+-}$ for $F_s=0$ and $p^{-+}<p^{+-}$ for $F_s<0$. 
To obtain the final condition required to determine $p^{-+}$ and $p^{+-}$ uniquely, we introduce the characteristic time $\tau_s$ over which individual expectations change due to random disturbances. Since $p^{-+}$ and $p^{+-}$ are per unit time, $\tau_s$ represents the characteristic time over which the total probability for a manager to reverse her expectation is unity:\footnote{
	Consider the impact of a random disturbance. At its end, the agent's state will remain or change to its opposite. Introduce $p^{'++}$ and $p^{'+-}$ as the probabilities for the agent, in state $+1$, to end up in state $+1$ or $-1$, respectively, and $p^{'--}$ and $p^{'-+}$ for the agent, in state $-1$, to end up in state $-1$ or $+1$, respectively. Thus, $p^{'++}+p^{'+-}=1$ and $p^{'--}+p^{'-+}=1$. Assuming that the ending state does not depend on the initial state, i.e. $p^{'++}=p^{'-+}$ and $p^{'--}=p^{'+-}$, we obtain $p^{'-+}+p^{'+-}=1$. If the disturbances are frequent on the timescale of interest, we can transform the discrete probabilities into continuous transition rates via $p^{-+}={p^{'-+}}/{\tau_s}$ and $p^{+-}={p^{'+-}}/{\tau_s}$ to recover equation \eqref{eq_transition_unity}.
}
\begin{equation}\label{eq_transition_unity}
(p^{-+}+p^{+-})\tau_s=1.
\end{equation}

Together conditions \eqref{eq_force} and \eqref{eq_transition_unity} imply the transition rates:
\begin{align}\label{eq_transition_rates}
p^{-+} &= \frac{1}{\tau_s\left(1+e^{-\alpha F_s}\right)}\; , & p^{+-} &= \frac{1}{\tau_s\left(1+e^{\alpha F_s}\right)}\; .
\end{align}
Equations \eqref{eq_transition_rates} allow us to rewrite \eqref{eq_s_rates} as
\begin{equation}\label{eq_s_macro}
\tau_s\dot{s} = -s + \tanh\left(F_s\right) = -s + \tanh\left(\beta_1 s + \beta_2 h + E_s\right),
\end{equation}
where ${\alpha}/{2}$ is absorbed into $\beta_1$ and $\beta_2$ without loss of generality. Note that $\tau_s$ acquires a dual meaning: at the micro level, $\tau_s$ is akin to the manager's average memory timespan; at the macro level, $\tau_s$ is the characteristic time of variation in the aggregate expectation of managers.

Applying this approach to model the dynamics of analyst expectations yields the same form of the evolution equation for information $h$: 
\begin{equation}\label{eq_h_macro}
\tau_h\dot{h} = -h + \tanh\left(F_h\right) = -h + \tanh\left(\beta_3 s + \beta_4 h + E_h\right),
\end{equation}
where $\tau_h$ represents the analyst's average memory timespan on the micro level and the characteristic time of the variation in the aggregate expectation of analysts on the macro level. Similarly, $F_h$ is the peer influence acting on the analysts' expectations, which is linear in $s$ and $h$ with sensitivities $\beta_3$ and $\beta_4$, and $E_h$ denotes general exogenous influences.

Equations \eqref{eq_s_macro} and \eqref{eq_h_macro} describe a generalized interactions-based process of decision-making. We now make several assumptions to adapt it to the capital demand discovery mechanism of the Dynamic Solow model (Figure \ref{fig_solow_flowchart}). 

First, we assume managers receive information only via analysts and accordingly set $E_s=0$. Second, we assume analysts are affected, first and foremost, by the news about economic development and only thereafter by all other news. More specifically, we assume the average analyst projects the output trend forward in time (extrapolative beliefs) and we treat all other relevant news as exogenous noise. Thus we set \begin{equation}E_h=\gamma\dot{y} + \xi_t,
\end{equation}
with sensitivity $\gamma$ and news noise $\xi_t$ acting on the timescale $\tau_\xi\ll\tau_h$. The latter implies that changes to expectations are impacted by short-term shocks with no relation to economic fundamentals (as suggested, for example, by \citet{AngeletosEtAl2020}).

Third, we establish separate timescales for information processing and expectation formation. That is, we assume information is received and processed much faster than it takes managers to adapt their long-term outlook and form investment decisions. Therefore: $\tau_h\ll\tau_s$. Fourth, as $\tau_h$ is much shorter than $\tau_s$, we assume direct interactions are less important for analysts than for managers and we take $\beta_3=\beta_4=0$ for simplicity.

The final step is to model the link between sentiment and capital demand. Consider a firm whose managers have just decided on capital allocation in line with their collective sentiment. The following day, all else being equal, the managers will not revisit this decision unless their sentiment changes. Therefore, in the short run where $t\ll\tau_s$ (that is, over time horizons where the memory of past sentiment persists), capital demand must be driven by change in sentiment. Conversely, over longer horizons where $t\gg\tau_s$, the connection between previous decisions and sentiment becomes weaker and, therefore, investment decisions must be based on the level of sentiment itself in the long run. For lack of simpler alternatives, we superpose these two asymptotic regimes, $\dot{k}_d\sim\dot{s}$ for $t\ll\tau_s$ and $\dot{k}_d\sim s$ for $t\gg\tau_s$, and, as a result, arrive at a complete system of equations for capital demand:
\begin{eqnarray}\label{eq_demand_sys}
    \dot{k}_d &=& c_1\dot{s} + c_2 s \label{eq_demand_kd},\\
    \tau_s\dot{s} &=& -s + \tanh\left(\beta_1 s + \beta_2 h\right), \label{eq_demand_s}\\
    \tau_h\dot{h} &=& -h + \tanh\left(\gamma\dot{y} + \xi_t\right),\label{eq_demand_h}
\end{eqnarray}
where $c_1>0$ and $c_2>0$ represent the capital demand sensitivity to a change in sentiment $\dot{s}$ and the level of sentiment $s$, respectively; and $\gamma>0$ represents the sensitivity of information $h$ to the state of the economy or, in other words, the strength of economic feedback.\footnote{\citet{GusevEtAl2015} derived equations \eqref{eq_demand_s} and \eqref{eq_demand_h} in the context of their generalized Ising model. The derivation presented here provides a more intuitive, albeit less rigorous, treatment. Equation \eqref{eq_demand_s}, with $h$ as an exogenous variable, was originally obtained by \citet{SuzukiKubo1968} for the classic \citet{Ising1925} model in statistical mechanics. Following  \citet{HaagWeidlich1983}, equation \eqref{eq_demand_s} (often in its stationary form) has frequently appeared in the socioeconomic context. For reviews of the interactions-based approaches reliant on the statistical mechanics methods and, particularly, the applications of the Ising model and its variants to the opinion dynamics problems in economics and finance, we recommend \citet{BrockDurlauf2001c}, \citet{Bouchaud2013} and \citet{Slanina2013}. We also note that, unlike equations \eqref{eq_demand_s} and \eqref{eq_demand_h} that are derived from the micro level, equation \eqref{eq_demand_kd} is obtained as a phenomenological relation between business sentiment and capital demand. Equation \eqref{eq_demand_kd} was initially suggested as a link between investor sentiment and stock market returns in \citet{GusevEtAl2015}.}

\subsection{Capital Market Clearing}\label{sec_model_capmkt}
At the relatively short time horizons relevant to information gathering, investment decision-making and production adjustment, prices are not flexible enough to efficiently match capital demand $k_d$ and supply $k_s$, which are determined independently from each other. Accordingly, we introduce an inelastic market clearing condition for log invested capital $k$ as
\begin{equation}\label{eq_capmkt_clearing}
    k = \min\left(k_s,k_d\right),
\end{equation}
to be satisfied at each moment in time. In contrast to the classic framework, in which all household savings are used in production, this condition implies that only a portion of savings will be invested should demand fall short of supply (with the remainder retained in household savings accounts).

Equation \eqref{eq_capmkt_clearing} is a local clearing condition that reflects the short-term price rigidity. Therefore, as was discussed in Section \ref{sec_model_structure}, this equation cannot remain valid over long-term horizons during which prices become sufficiently flexible to match demand and supply. As such, we supplement \eqref{eq_capmkt_clearing} with an asymptotic clearing condition that holds in the timescale of long-term economic growth:\footnote{The asymptotic relation \eqref{eq_capmkt_lt} means that the relative error between $k_s$ and $k_d$ goes to zero as $t$ goes to infinity.}
\begin{equation}\label{eq_capmkt_lt}
	k_s \sim k_d \quad \text{for} \quad t\geq O\left(1/\varepsilon\right) \gg 1.
\end{equation}
Together, equations \eqref{eq_capmkt_clearing} and \eqref{eq_capmkt_lt} interlink the supply and demand components and close the Dynamic Solow model. 

At this point, it may be useful to discuss the characteristic timescales in the model. The timescales we have encountered are differentiated in length such that $\tau_\xi\ll\tau_h\ll\tau_s\ll\tau_y\ll1/\varepsilon$. 
Economically, information gathering occurs on a relatively short timescale, $\tau_h$ (with the publication of, for example, monthly and quarterly corporate reports and industry data releases); investment decisions require more time, $\tau_s$ (as processed through, for example, annual board meetings); and the implementation of changes to production levels takes much longer, $\tau_y$ (the time needed for material adjustments such as infrastructure development). We set $\tau_h=25$, $\tau_s=250$ and $\tau_y=1000$ in units of business days (250 business days = 1 year). We further assume the timespan of exogenous news events to be on average one week and set $\tau_\xi=5$ as the news noise decorrelation timescale (see \ref{appx_parameters}). Finally, we take technology growth rate $\varepsilon=2.5\times10^{-5}$, which implies the timescale of 160 years.\footnote{
	In equation \eqref{eq_cd_basic}, the term $\exp(\varepsilon t)$, commonly referred to as total factor productivity, yields a growth rate of about 0.6\% p.a. for $\varepsilon=2.5\times10^{-5}$, which is not far from the estimates based on the 2005-2016 period using \cite{Fernald2016}.
}

\section{Two Limiting Cases}\label{sec_limitcases}
In this section, we inspect two cases that follow from the market clearing condition \eqref{eq_capmkt_clearing}: first, the supply-driven case, $k_d>k_s$ such that $k=k_s$, which recovers a Solow-type growth economy; and, second, the demand-driven case, $k_d<k_s$ such that $k=k_d$, in which the economic fluctuations emerge.

\subsection{Supply-Driven Case $k_d>k_s$}\label{sec_ks_limit}
In the supply-driven case, the market clearing condition yields $K=K_s$ (firms use all available capital for production)\footnote{It is convenient to use the non-logarithmic variables for this analysis.}. Consequently, the Dynamic Solow model is reduced to equations \eqref{eq_y_dot_first} and \eqref{eq_cap_sup}
which can be expressed as a single second-order differential equation:
\begin{equation}\label{eq_limit_ks_second_order}
    \tau_Y\ddot{K} + (1+\tau_Y\delta)\dot{K} + \delta K = \lambda K^\rho e^{\varepsilon t}.
\end{equation}

\begin{figure}[H]
    \centering
    \includegraphics[width=0.8\linewidth]{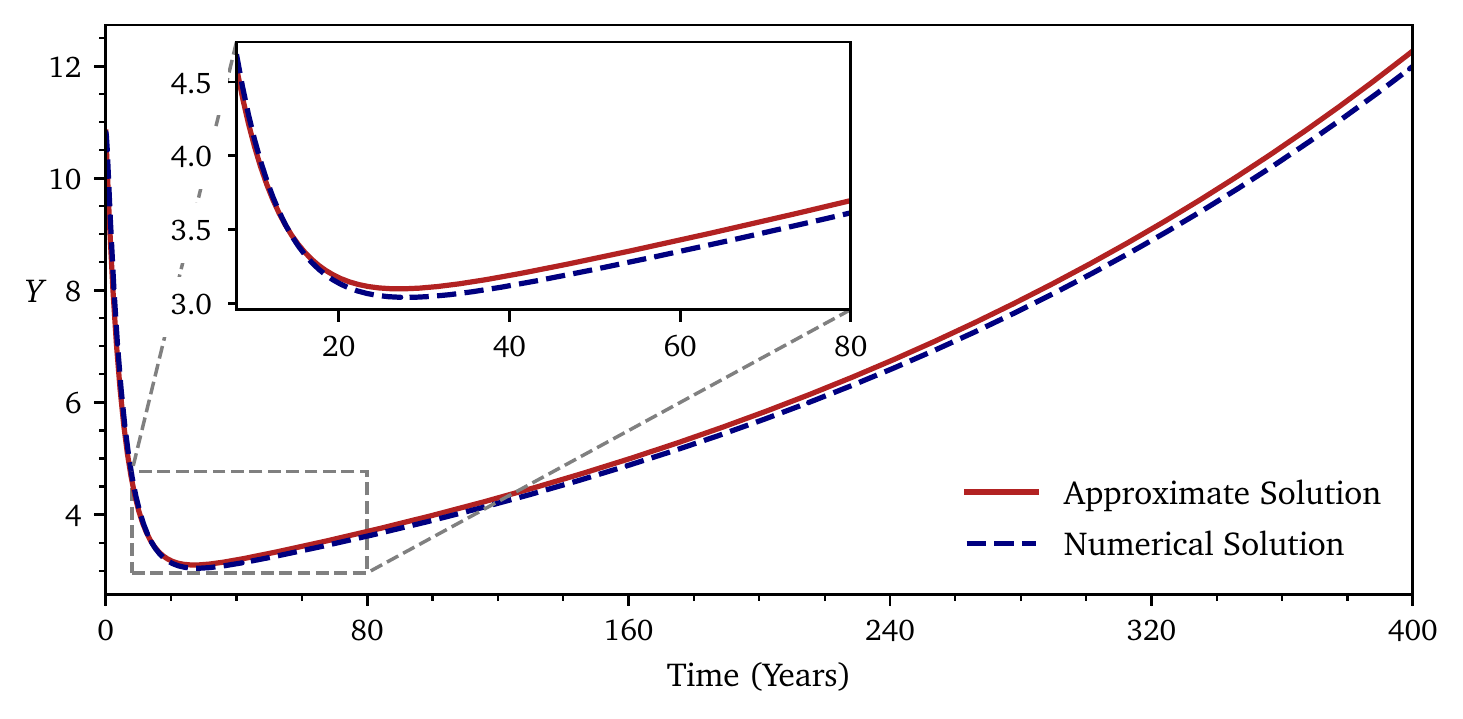}
    \caption{Output $Y(t)$ in the supply-driven case represented by the numerical solution of equation \eqref{eq_limit_ks_second_order} (dashed blue line) and the approximate solution \eqref{eq_bla} (solid red line). The precision of the approximate solution is improved with a greater timescale separation $\tau_y\ll1/\varepsilon$. Parameters: $\rho=1/3$, $\tau_y=1000$, $\lambda=0.15$, $\varepsilon=10^{-5}$ and $\delta=0.02$ with an integration constant of $B=1.5$. The inset box highlights the intermediate adjustment of output from an arbitrary initial value to the equilibrium growth path.}
    \label{fig_limitks}
\end{figure}

For $t\sim1/\varepsilon$ and longer time intervals, the derivative terms in equation \eqref{eq_limit_ks_second_order} become negligibly small and we recover the equilibrium growth path. On shorter timescales, $t\sim\tau_y$, equation \eqref{eq_limit_ks_second_order} describes adjustment towards the equilibrium growth path. These two effects can be observed simultaneously by deriving an approximate solution to equation \eqref{eq_limit_ks_second_order} for $t\geq O(\tau_y)$ (see \ref{appx_boundarylayer}). The resulting production path is given by
\begin{equation}\label{eq_bla}
    Y = \left(\frac{\lambda}{\delta}\right)^{\frac{\rho}{1-\rho}} \left(\left(Be^{-\left(\frac{1-\rho}{\tau_y}\right)t}+1\right)^{\frac{1}{1-\rho}} +e^{\left(\frac{\varepsilon}{1-\rho}\right)t}-1\right),
\end{equation}
where $B$ is the constant of integration.\footnote{
	This derivation is valid for the parameter values provided in \ref{appx_notation}, subject to the simplifying assumption $\tau_y\delta\gg1$ which allows us to obtain the solution in a compact form.
} Equation \eqref{eq_bla} explains the output dynamics between intermediate and long-term timescales, capturing both the long-term growth of the classic Solow model (given by the second exponent) and the intermediate relaxation towards the same (given by the first exponent). The approximate analytic solution \eqref{eq_bla} and the exact numerical solution to equation \eqref{eq_limit_ks_second_order} are compared in Figure \ref{fig_limitks}.

\subsection{Demand-Driven Case $k_d<k_s$}\label{sec_kd_limit}
In the demand-driven case, the market clearing condition yields $k=k_d$. The Dynamic Solow model is specified at this limit by equations \eqref{eq_y_dot} and \eqref{eq_demand_kd}-\eqref{eq_demand_h} (in this case, equation \ref{eq_ks_dot} decouples and no longer affects production). To facilitate our analysis, we introduce the variable $z=\rho k_d + \varepsilon t - y$, which makes the model solutions bounded in the $(s,h,z)$-space (see \ref{appx_asymp_derivation}). Economically, $z$ represents the direction and strength of economic growth. This follows from rewriting equation \eqref{eq_y_dot} as $\tau_y\dot{y}=e^z-1$, noting that for $z>0$ production expands, for $z<0$ it contracts and $z=0$ is a production fixed point. Using $z$, we re-express the model as a three-dimensional dynamical system that is bounded and autonomous in the absence of exogenous noise $\xi_t$:
\begin{subequations}\label{eq_kd_shz_system}
\begin{align}
\dot{z} &= \rho c_1 \dot{s} + \rho c_2 s - \omega_Y \left(e^{z}-1\right)+\varepsilon\label{eq_z_z_dot}\\
\tau_s \dot{s} &= -s + \tanh\left(\beta_1 s + \beta_2 h \right)\label{eq_z_s_dot}\\
\tau_h \dot{h} &= -h + \tanh\left(\gamma\omega_y\left(e^{z}-1\right) +\xi_t \right)\label{eq_z_h_dot},
\end{align}
\end{subequations}
where, for convenience, $\omega_y = 1/\tau_y$. 

This dynamical system is parametrized and examined in detail in \ref{appx_parameters}. For the relevant range of parameters it has three equilibria: a stable focus where sentiment is positive ($s>0$) and the economy is expanding ($z>0$), a stable focus where sentiment is negative ($s<0$) and the economy is contracting ($z<0$) and an unstable saddle point in between.\footnote{For convenience, we classify 3D equilibrium points using more familiar 2D terminology. As such: (i) the stable (unstable) node has three negative (positive) real eigenvalues; (ii) the focus has one real and two complex eigenvalues and is stable if the real eigenvalue and the real parts of complex eigenvalues are all negative and unstable otherwise; and (iii) the saddle is always unstable as it has three real eigenvalues that do not all have the same sign. In the figures, the stable points are green and unstable points are red, while the nodes are marked by triangles, foci by squares and saddles by circles.\label{point notatrion}} The location, basin of attraction and stability of the equilibria are primarily affected by the parameters $c_2$ (sensitivity to sentiment levels) and $\gamma$ (sensitivity to economic feedback). In particular, an  increasing $c_2$ strengthens convergence towards the equilibria, so the system acquires greater stability.
\begin{figure}
	\centering
\begin{subfigure}[b]{0.95\textwidth} 
    \centering
	\includegraphics[trim=0 7 0 60, clip, width=0.65\textwidth]{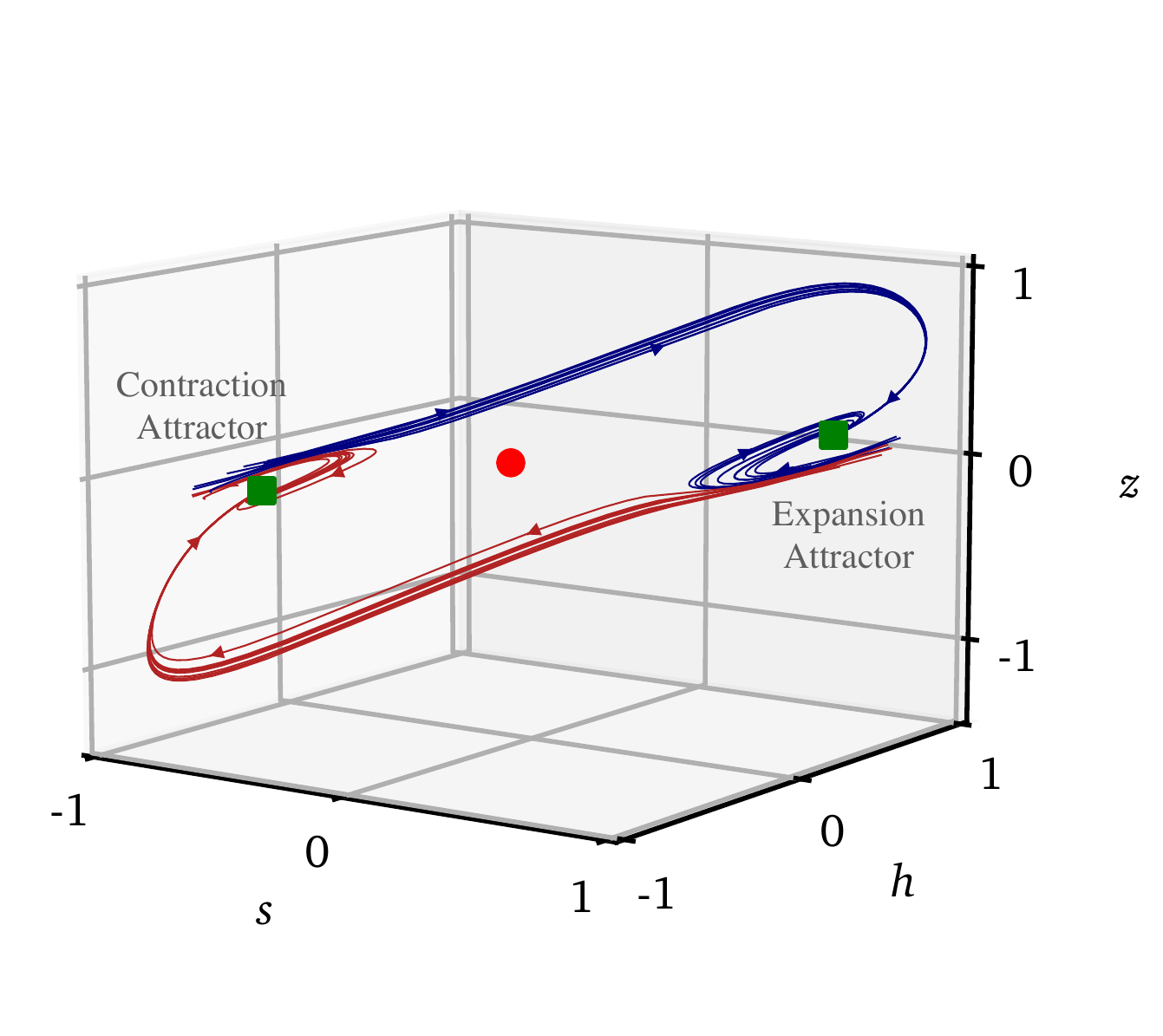}
	\caption{Coherence resonance for $c_2=7\times10^{-4}$ and $\gamma=2000$. This subcritical regime presents a bi-stable configuration of equilibria: green squares denote the two stable foci and the red circle an unstable saddle. Red trajectories terminate at the $s<0$ focus in which the economy contracts and blue trajectories terminate at the $s>0$ focus in which the economy expands. The long trajectories passing near one focus and ending at the other are of a particular interest as they provide the pathway for the economy's regime transitions.}
	\label{fig_demand_limit_base_phase_3d}
\end{subfigure}

\begin{subfigure}[b]{0.95\textwidth} 
    \centering 
	\includegraphics[trim=0 7 0 60, clip, width=0.65\textwidth]{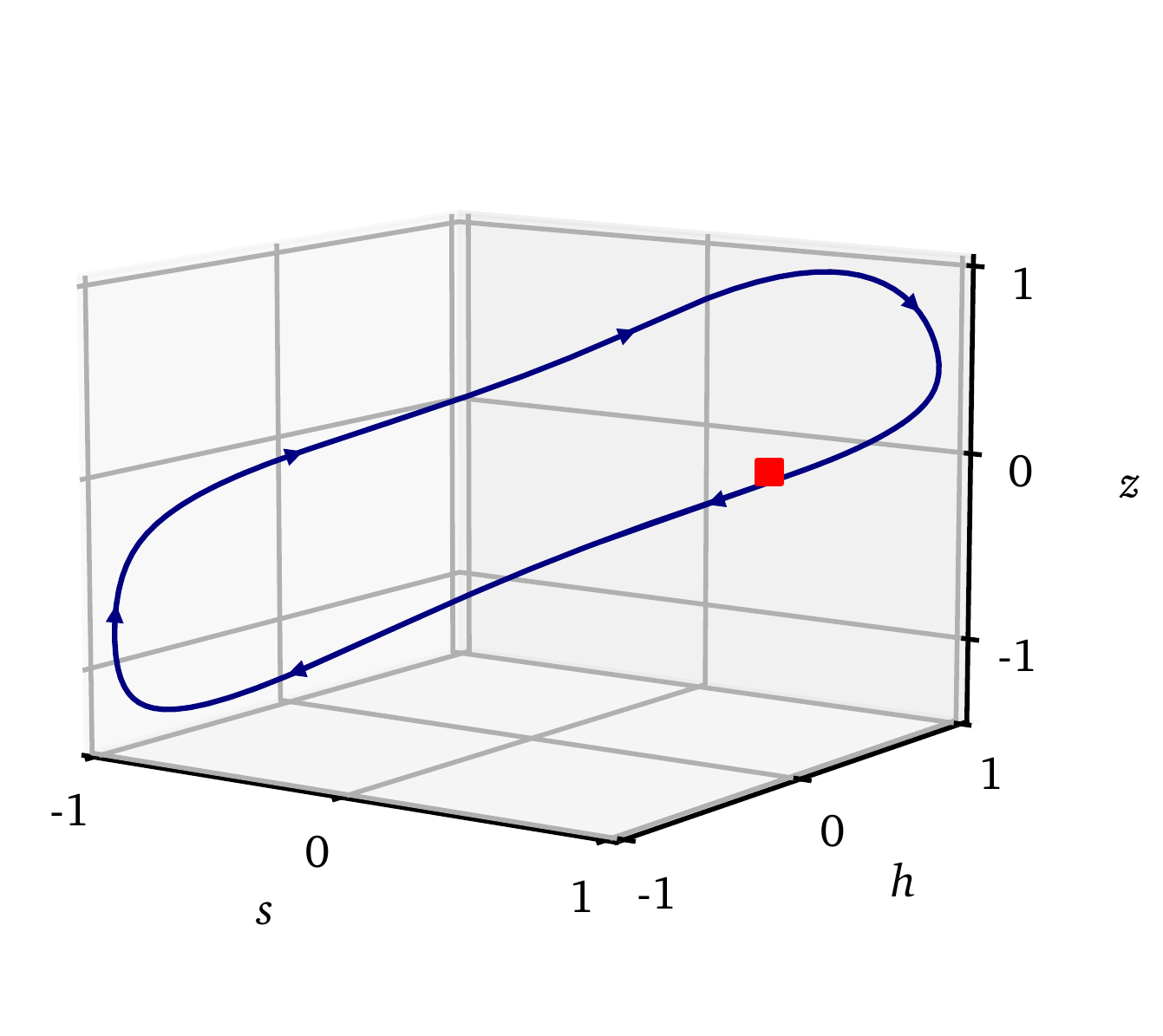}
	\caption{Limit cycle for $c_2=1\times10^{-4}$ and $\gamma=4000$. In this supercritical regime, only the positive ($s>0$) equilibrium point survives, having bifurcated into an unstable focus, and a large stable limit cycle emerges that propels the economy between contraction and expansion with a constant frequency.}
	\label{fig_demand_limit_cycle_phase_3d}
\end{subfigure}

\caption{3D phase portraits ($\xi_t=0$) in the $(s,h,z)$-space: (a) the coherence resonance regime and (b) the limit cycle regime. The long trajectories in (a) can be viewed as segments of the limit cycle in (b), which remain unconnected for $\xi_t=0$. Parameters other than $c_2$ and $\gamma$ are from the base case in \ref{appx_notation}.}
\label{fig_demand_two_phases_3d}
\end{figure}

\begin{figure}
	\centering 
	\includegraphics[width=1.0\textwidth]{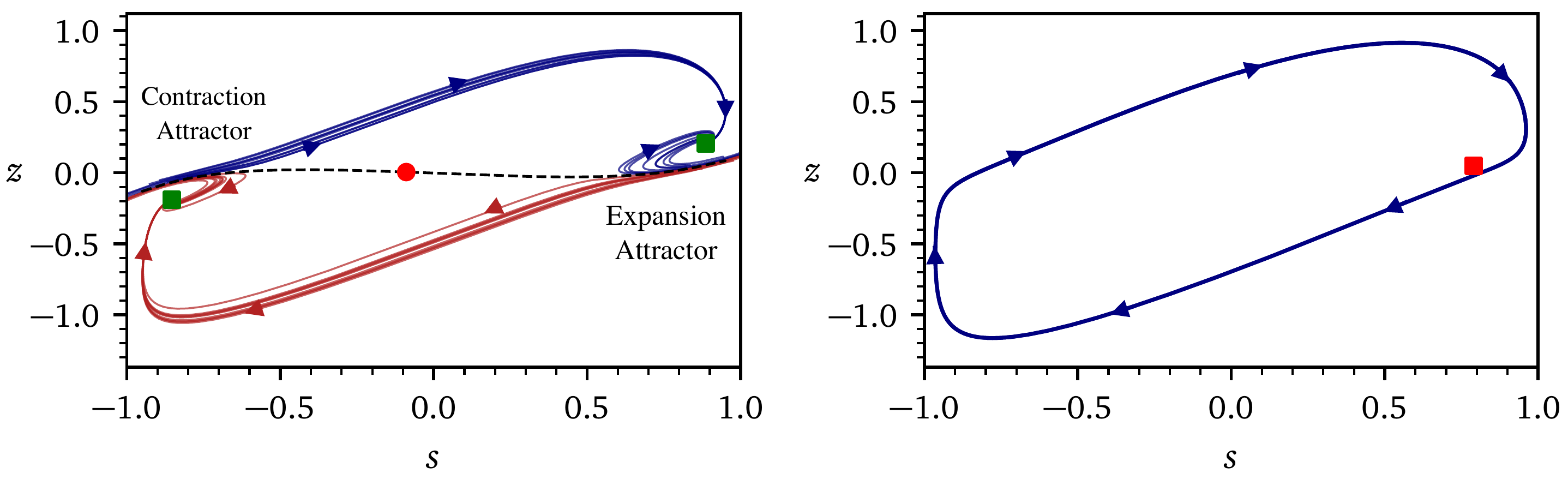}
	\caption{Left: The coherence resonance phase portrait in Figure \ref{fig_demand_limit_base_phase_3d} projected on the $s,z$-plane. The approximate boundary between the contraction and expansion regions is indicated by a dashed black line. When the economy moves along the long trajectories, it traverses quickly the distance in $s$ between the foci. However, the movement speed declines sharply with the proximity to equilibrium. As a result, the economy slowly ascends along the $z$-axis towards the left focus in the contraction mode and slowly descends to the right focus in the expansion mode. As the economy is undergoing this gradual drift, the distance to the boundary separating contraction and expansion diminishes. Right: The limit cycle phase portrait in Figure \ref{fig_demand_limit_cycle_phase_3d} projected on the $s,z$-plane.} 
	\label{fig_demand_limit_sz_comparison}
\end{figure}

If $c_2$ is below a certain critical value, equations \eqref{eq_kd_shz_system} generate a periodic limit cycle. The idea that limit cycles provide a mechanism of economic fluctuations dates back to \citet{Kalecki1937}, \citet{Kaldor1940}, \citet{Hicks1950} and \citet{Goodwin1951}. The empirical irrelevance of periodic limit cycles led to a diminished interest in this research direction\footnote{
	 Nevertheless, a similar line of research has been pursued in overlapping generations models and innovation cycles. See \citet{Hommes2013} and \citet{BeaudryEtAl2020} for references.
}; however, \citet{BeaudryEtAl2020} have reinitiated the discussion by proposing that cyclicality can arise from stochastic limit cycles ``wherein the system is buffeted by exogenous shocks, but where the deterministic part of the system admits a limit cycle''. In our system, exogenous news noise $\xi_t$ similarly detunes limit cycle periodicity. This mechanism, however, cannot explain the ``small shock, large crisis'' effect or reproduce the general variability present in the real-world economy. At the other extreme, our system generates noise-prevailing behaviors with weak cyclicality. Neither extreme accurately reflects empirical observations and thus we seek a sensible balance between these features in a parameter regime that produces significant dynamic effects but precedes the limit cycle formation (\ref{appx_parameters}). 

\begin{figure}[htb!]
	\centering
	\includegraphics[trim=0 0 0 15, clip, width=0.8\textwidth]{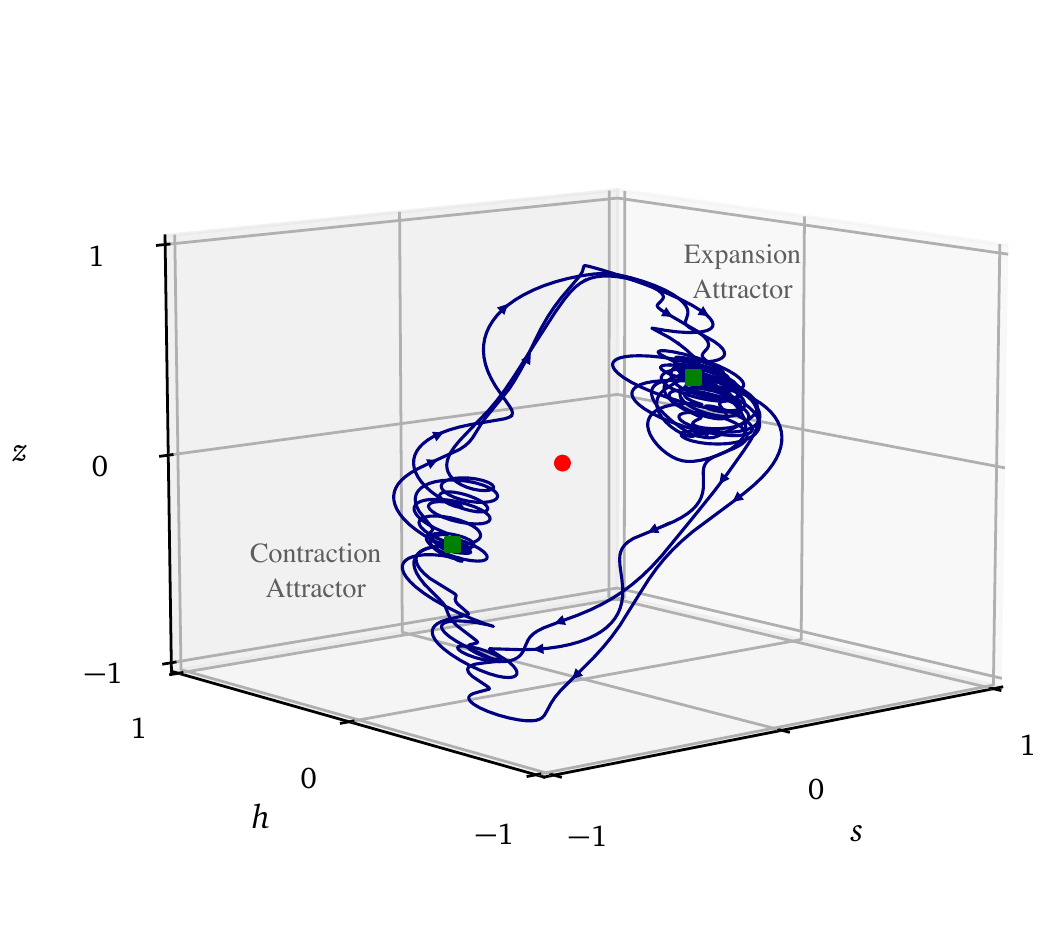}
	\caption{A simulated economy's path for nonzero $\xi_t$ in the coherence resonance case, smoothed by a Fourier filter to remove harmonics with periods less than 500 business days for a better visualization in the 3D phase space. The regions where the trajectory is dense indicate the contraction and expansion attractors, around which the economy spends most of its time. The relatively straight path segments between the attractors correspond to the economic regime transitions that occur on a relatively rapid timescale. Parameters are from the base case in \ref{appx_notation}.}
	\label{fig_demand_limit_base_series_3d}
\end{figure}

To this end, we consider a subcritical regime with $c_2$ above but close to its critical value at which the limit cycle emerges. In this situation, henceforth referred to as the coherence resonance regime, the foci are always stable, thus acting as attractors entrapping the economy. In Figures \ref{fig_demand_two_phases_3d} and \ref{fig_demand_limit_sz_comparison},
we compare the phase portraits (i.e. the system trajectories for $\xi_t=0$) of the coherence resonance and limit cycle regimes. In the coherence resonance case, we take note of the unclosed largescale trajectories that pass near one attractor and converge to the other. These trajectories, which can be viewed as segments of a limit cycle, are the pathways along which the economy moves between contraction and expansion.

\begin{figure}[htb!]
	\centering
	\includegraphics[width=1.0\textwidth]{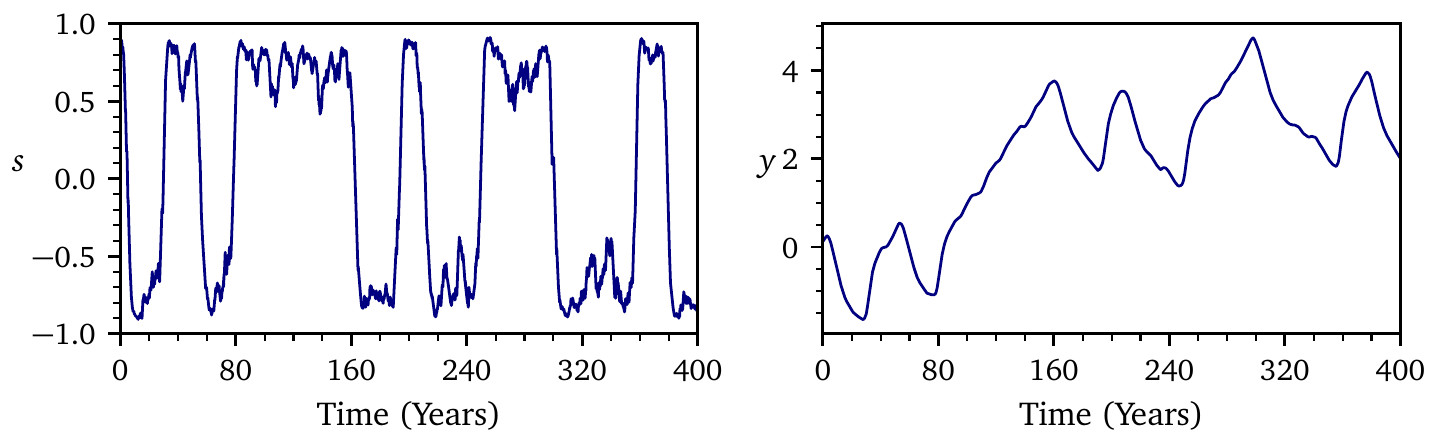}
	\caption{Sentiment $s(t)$ (left) and output $y(t)$ (right) on a simulated economy's path for nonzero $\xi_t$ in the coherence resonance regime. Output undergoes fluctuations within a general growth trend as sentiment evolution exhibits a distinct bi-stable pattern. Parameters are from the base case in \ref{appx_notation}.}
	\label{fig_demand_limit_base_series_sy}
\end{figure}

The dynamics of business cycles for nonzero $\xi_t$ are visualized in Figure \ref{fig_demand_limit_base_series_3d}. The economy's trajectory displays distinctly bi-stable behavior as it spends most of its time near each focus and transits swiftly between them. When captive to an attractor, the trajectory follows an orbit around the corresponding focus, buffeted by exogenous noise $\xi_t$, preventing it from settling. Simultaneously, the economy drifts slowly towards the boundary between attracting regions (Figure \ref{fig_demand_limit_sz_comparison}(left)),
 making it easier for a random news shock to thrust it across the boundary to be caught by the other attractor. The news shocks $\xi_t$ thus fulfill a dual purpose: they perturb the economy from equilibrium and provide a trigger that alternates the economic regime between expansions and recessions.  

This mechanism can be classified as coherence resonance, a phenomenon whereby noise applied to a dynamical system leads to a quasiperiodic response \citep{PikovskyKurths1997}. Coherence resonance normally occurs in bi-stable systems that are stochastically forced and in which key variables evolve on different timescales. The Dynamic Solow model satisfies these requirements: (i) news shocks provide a stochastic force; (ii) two stable equilibria emerge in the relevant parameter range; and (iii) the separation of characteristic timescales follows from the dynamics of corporate decision-making processes. The three-dimensionality of equations \eqref{eq_kd_shz_system} introduces an important novel feature into the classic two-dimensional case of coherence resonance: the above-mentioned slow drift of the economy's trajectory, which gradually increases the probability of regime transition.\footnote{
	This drift imposes a slow timescale $t\geq O(\tau_y)$ on the frequency of the economy's fluctuations by modulating the probability of regime transition. This effect was first observed and explained for a similar dynamical system in \citet{KroujilineEtAl2019}.
} This novel feature nonetheless leaves the basic mechanism unchanged: exogenous noise forces the economy across the boundary separating the regions of different dynamics, effectively reconnecting the trajectories between attractors. As a result, the economy undergoes quasiperiodic fluctuations consisting of alternating periods of expansion and recession punctuated by sharp transitions (as in Figure \ref{fig_demand_limit_base_series_sy}).

While both coherence resonance and limit cycle occur in the economically realistic range of parameters (see \ref{appx_parameters}), we will show that only coherence resonance is compatible with the long-term economic growth when studying the asymptotic convergence of capital supply and demand in the presence of fluctuations driven by the coherence resonance and limit cycle mechanisms in Section \ref{sec_results_asymp}. 


\section{Business Cycles and Long-Term Growth in the General Case}\label{sec_results}
While the supply- and demand-driven cases have been instructive for highlighting the mechanisms underlying economic dynamics, their applicability as standalone models is limited as supply and demand converge in the long run (equation \eqref{eq_capmkt_lt}). As such, our primary focus is on the general case in which supply and demand coevolve, potentially leading to an interplay of supply- and demand-driven dynamics. 
We formulate the general case in Section \ref{sec_results_formulation}, study long-term growth rates in Section \ref{sec_results_asymp} and examine economic fluctuations in Section \ref{sec_results_cycle}.

\subsection{Formulation of the General Case}\label{sec_results_formulation}
In the general case, invested capital $k$ can alternate between $k_d$ (demand-driven regime) and $k_s$ (supply-driven regime) in accordance with the market clearing condition \eqref{eq_capmkt_clearing}. As discussed in Section \ref{sec_model_structure}, firms' decision-making processes are influenced by feedback from the economy. However, the supply-driven regime represents a special situation in which firms' investment decisions do not affect economic output as production is determined in this case solely by capital availability. In other words, the supply-driven regime implies a Solow-type growth economy propelled by expectations of future consumption so high as to induce firms to utilize all capital supplied by households in production. Therefore, $\dot{y}$, which is positive in this regime, holds no additional information for managers, who are already overwhelmingly bullish about the economy. The idiosyncratic news $\xi_t$ remains the only source of nontrivial information, thereby becoming the focus of managers and analysts alike. Thus, economic feedback $\gamma\dot{y}$ vanishes as a decision factor in the supply-driven regime.

Following this argument, we account for regime-dependent variation in feedback strength by introducing a regime-specific factor $H(k_d, k_s)$ that regulates the impact of feedback in equation \eqref{eq_demand_h}:
\begin{eqnarray}
	\tau_h \dot{h} &=& -h + \tanh\left(\gamma\dot{y}H(k_s,k_d) +\xi_t\right)\label{eq_h_dot},
\end{eqnarray}
where
\begin{equation}\label{eq_H_switch}
	H(k_s,k_d) = 
	\begin{cases}
		1 &\text{if }k_d\leq k_s\\
		0 &\text{if }k_d>k_s
	\end{cases} .
\end{equation}

The Dynamic Solow model is then represented in the general case by the following system of equations:
\begin{align}
	\tau_Y \dot{y} &= e^{\rho k + \varepsilon t - y} - 1 \label{eq_res_full_y_dot},\\
	\dot{k}_s &= \lambda e^{y-k_s} - \delta e^{k-k_s}\label{eq_res_full_ks_dot},\\
	\dot{k}_d &= c_1 \dot{s} + c_2 s \label{eq_res_full_kd_dot},\\
	\tau_s \dot{s} &= -s + \tanh\left(\beta_1 s + \beta_2 h \right)\label{eq_res_full_s_dot},\\
	\tau_h \dot{h} &= -h + \tanh\left(\gamma\dot{y}H(k_s,k_d) +\xi_t\right)\label{eq_res_full_h_dot}, \\
	k &= \min(k_d, k_s) \label{eq_res_full_clearing}, \\
	k_s &\sim k_d \quad \text{for} \quad t\geq O(1/\varepsilon)\gg1,\label{eq_res_full_matching}
\end{align}
where \eqref{eq_res_full_y_dot} is the dynamic equation governing production; \eqref{eq_res_full_ks_dot} describes the motion of capital supply; \eqref{eq_res_full_kd_dot}-\eqref{eq_res_full_h_dot} govern the feedback-driven dynamics that link information $h$, sentiment $s$ and capital demand $k_d$; \eqref{eq_res_full_clearing} defines the locally-inelastic market clearing condition; and \eqref{eq_res_full_matching} represents long-term market clearing that takes the form of an asymptotic boundary condition at large $t$. 

\subsection{Growth and Convergence in the Long Run}\label{sec_results_asymp}



The Dynamic Solow model \eqref{eq_res_full_y_dot}-\eqref{eq_res_full_matching} covers two regimes with different dynamics: a demand-driven regime with endogenous fluctuations and a supply-driven regime without them. Both regimes are expected to participate in the model's general case, owing to the convergence of supply and demand in the long run under equation \eqref{eq_res_full_matching}. 

Equation \eqref{eq_res_full_matching} is central to our present analysis. Based on the regime definitions, this equation is satisfied when supply grows faster than demand in the supply-driven regime and, conversely, when demand grows faster than supply in the demand-driven regime. Under the demand-driven regime, the two possible mechanisms of fluctuations -- limit cycle and coherence resonance -- may entail different growth rates, validating the mechanism if demand grows fast enough to satisfy \eqref{eq_res_full_matching} and invalidating it otherwise. 

This section aims to determine (i) the impact of fluctuations on growth; (ii) the mechanism of fluctuations compatible with equation \eqref{eq_res_full_matching}; and (iii) the actual growth dynamics realized in the model. We first consider separately the supply- and demand-driven regimes (Sections \ref{sec_results_asymp_sc} and \ref{sec_results_asymp_dc}) and then tackle the general case (Section \ref{sec_results_asymp_gc}). \ref{appx_asymp_derivation} provides the derivations of the equations herein.

\subsubsection{Asymptotic Growth in the Supply-Driven Case ($k_d>k_s$)}\label{sec_results_asymp_sc}

\begin{figure}
    \centering
    \includegraphics{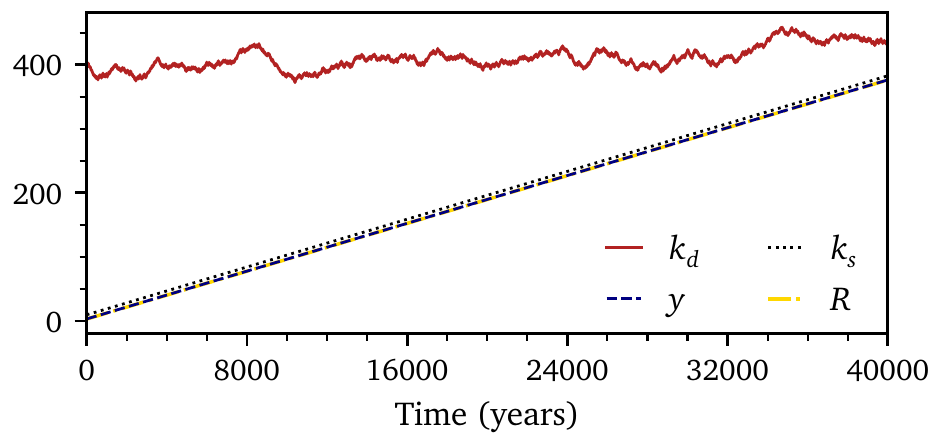}
    \caption{A long-term simulation that captures asymptotic growth in the enforced supply-driven regime ($k=k_s$). Output $y$ and supply $k_s$ grow steadily at the Solow rate $R$ while demand $k_d$ stagnates. As such, $k_s$ will eventually reach and exceed $k_d$, at which point the supply-driven regime will be succeeded by the demand-driven regime. Parameters are from the base case in \ref{appx_notation}.}
    \label{fig_results_asymp_supply}
\end{figure}

We show in \ref{appx_asymp_derivation} that the economy's long-term growth in the supply-driven case is given by
\begin{align}\label{eq_asympotitc_prod_supply}
	y_0 &= k_{s0} = \frac{\varepsilon}{1-\rho} \equiv R,\\
    k_{d0} &= 0,
\end{align}
where $y_0$, $k_{s0}$ and $k_{d0}$ represent, respectively, the log output, log supply and log demand growth rates; $\rho=1/3$ is the capital share in production; and $R$ denotes the classic Solow growth rate. As expected, the growth rate is not influenced by demand dynamics and matches $R$. These estimates are verified by numerical simulations (see Figure \ref{fig_results_asymp_supply}). Note that supply always catches up with demand as $k_{s0}>k_{d0}$ in this case.

\subsubsection{Asymptotic Growth in the Demand-Driven Case ($k_d<k_s$)}\label{sec_results_asymp_dc}

We show in \ref{appx_asymp_derivation} that the economy's long-term growth in the demand-driven case satisfies 
\begin{equation} \label{asymptotic relation}
y_0 = k_{s0} = R + \rho\left(k_{d0} - R\right).
\end{equation}
According to this equation, output becomes dependent on demand, which itself undergoes endogenous fluctuations, and thus demand dynamics affect the long-term behavior of the economy. The magnitude of $k_{d0}$ determines whether the economy expands faster or slower than the classic Solow economy: $y_0>R$ if $k_{d0}>R$ and $y_0<R$ if $k_{d0}<R$ (the latter condition including an important case when $k_{d0}=0$ that yields an especially slow growth rate $y_0 = k_{s0} = \varepsilon$). Next, we estimate $k_{d0}$ numerically under the effect of limit cycle and coherence resonance.

\begin{figure}[htb]
    \centering
    \includegraphics{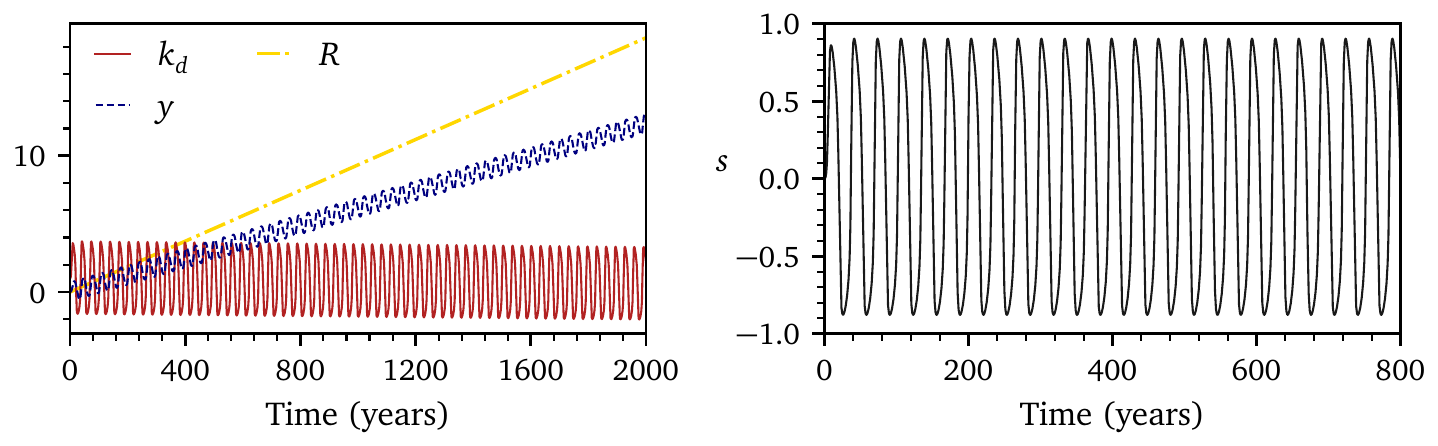}
    \caption{A long-term simulation that captures asymptotic growth in the enforced demand-driven regime ($k=k_d$) powered by a periodic limit cycle ($\xi_t=0$) with $\gamma=1000$, $c_2=2\times10^{-5}$ and all other parameters from the base case in \ref{appx_notation}. Left: Production $y$ grows at a rate lower than the Solow rate $R$ while demand $k_d$ stagnates (and, in fact, appears to gradually decrease, which could be attributed to the slight asymmetry of the limit cycle with respect to $s$). Since $k_s$ and $y$ grow at the same rate (equation \eqref{asymptotic relation}), $k_d$ cannot catch up with $k_s$. Right: Sentiment $s(t)$ demonstrates the limit cycle's periodicity.}
    \label{fig_results_asymp_purecycle}
\end{figure}

\begin{figure}[htb]
    \centering
    \includegraphics{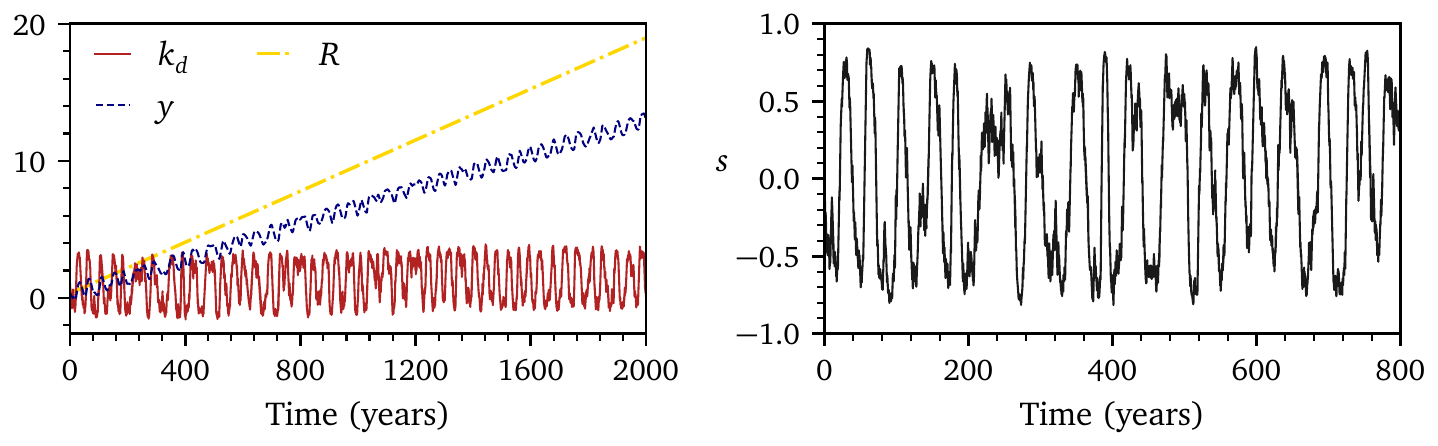}
    \caption{A long-term simulation that captures asymptotic growth in the enforced demand-driven regime ($k=k_d$) powered by a stochastic limit cycle ($\xi_t\neq0$) with the same parameters as in Figure \ref{fig_results_asymp_purecycle}. Left: Production $y$ grows at a rate lower than the Solow rate $R$ while capital demand $k_d$ stagnates (exogenous noise evidently erasing the limit cycle's asymmetry visible in Figure \ref{fig_results_asymp_purecycle}). Since $k_s$ and $y$ grow at the same rate (equation \eqref{asymptotic relation}), $k_d$ cannot catch up with $k_s$. Right: Sentiment $s(t)$ is no longer periodic due to the impact of $\xi_t$.}
    \label{fig_results_asymp_noisecycle}
\end{figure}

Figure \ref{fig_results_asymp_purecycle} depicts the growth dynamics driven by a periodic limit cycle ($\xi_t=0$). We observe that $k_{d0}$ stays close to zero and $y_0$ and $k_{s0}$ match $\varepsilon$ closely in accordance with \eqref{asymptotic relation}, meaning the economy grows only through improvements in production efficiency. Figure \ref{fig_results_asymp_noisecycle} displays similar dynamics for the limit cycle perturbed by exogenous noise $\xi_t$. It follows that limit cycles, whether periodic or stochastic, lead to a growth rate of less than $R$. 

The above result can be explained by noting that an economy on a limit cycle trajectory spends roughly an equal amount of time in expansion ($s>0$) as in contraction ($s<0$) and, consequently, $s$ exhibits on this trajectory a long-term average value of zero. In \ref{appx_asymp_derivation}, we find that $k_{d0}$ is proportional to the long-term average of $s$, implying $k_{d0}$ tends to zero as well; therefore, demand can never catch up with supply due to the difference in their growth rates. In sum, the fluctuations generated by a limit cycle detract from long-term growth and fail to satisfy equation \eqref{eq_res_full_matching}.


\begin{figure}[htb]
    \centering
    \includegraphics{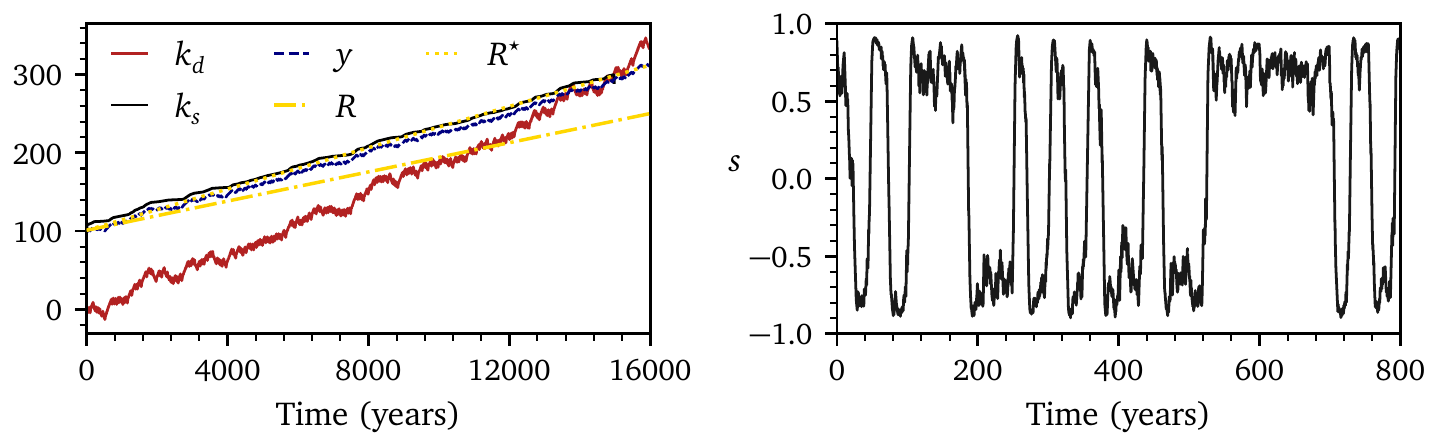}
    \caption{A long-term simulation that captures asymptotic growth in the enforced demand-driven regime ($k=k_d$) powered by coherence resonance in the base case parameter regime (\ref{appx_notation}). Left: Output $y$ and supply $k_s$ expand at the rate $R^\star>R$ and demand $k_d$ grows faster than both $y$ and $k_s$. As such, $k_d$ eventually reaches and exceeds $k_s$, at which point the demand-driven regime is succeeded by the supply-driven regime. Right: Sentiment $s(t)$ exhibits a bi-stable behavior typical of coherence resonance.}
    \label{fig_results_asymp_demand}
\end{figure}

Coherence resonance induces a drastically different long-term dynamic despite the visually similar fluctuations (see Figure \ref{fig_results_asymp_demand}). Demand grows asymptotically at $k_{d0}>R$, leading to accelerated economic growth of $y_0=k_{s0} \equiv R^\star>R$ in accordance with equation \eqref{asymptotic relation}. This fast-paced growth, made possible by excess capital, $k_s>k_d$, available for investment in the demand-driven regime, is explained by technological progress ($\varepsilon>0$), causing the economy to spend on average more time in expansion than contraction. We further observe that $k_{d0}>R^\star$; that is, demand grows faster than both supply and output.\footnote{This result is consistent with equation \eqref{asymptotic relation}, which can be rewritten as $k_{d0}-R^\star=((1-\rho)/\rho)(R^\star-R)$, so that $k_{d0}>R^\star$ since $R^\star>R$ and $\rho<1$.} Therefore, demand powered by coherence resonance always catches up with supply.

\subsubsection{Asymptotic Growth in the General Case}\label{sec_results_asymp_gc}

We have shown that fluctuations affect growth in the demand-driven regime of the Dynamic Solow model. In particular, limit cycles generate fluctuations that contribute negatively to growth, thus failing to satisfy the asymptotic boundary condition \eqref{eq_res_full_matching}. Therefore, such fluctuations cannot be realized, which rules out limit cycles as the mechanism responsible for business cycles.  

\begin{figure}[htb]
    \centering
    \includegraphics{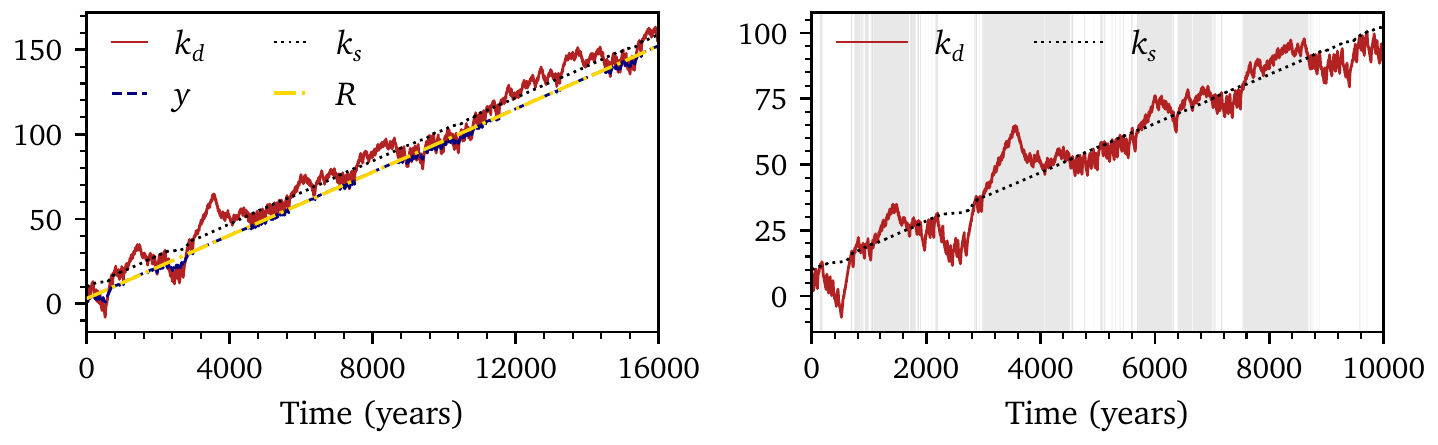}
    \caption{A long-term simulation that captures asymptotic growth in the general case. Left: Output $y$, supply $k_s$ and demand $k_d$ grow at the Solow rate $R$, demonstrating the asymptotic convergence on the economy's trajectory. Right: The interplay of supply- and demand-driven regimes on a subsection of the same trajectory. Shaded segments correspond to periods when $k_d>k_s$. Parameters are from the base case in \ref{appx_notation}.}
    \label{fig_results_asymp_general}
\end{figure}

By contrast, coherence resonance produces fluctuations that contribute positively to growth, so that demand always catches up with supply. As this occurs, the system transits into the supply-driven regime in which supply grows faster than demand. Once supply has exceeded demand, the system switches back into the demand-driven dynamics. The regime cycle has thus come full circle, ensuring \eqref{eq_res_full_matching} is satisfied in the long run. As such, the economy's path realized in the general case is forged by a regime interplay where the supply-driven equilibrium dynamics and the demand-driven fluctuations, powered by coherence resonance, continuously succeed one another. 

Our simulations show the economy grows asymptotically at the Solow rate $R$. This result is not entirely unexpected. As capital supply and demand converge over the long run, capital invested into production during the supply- and demand-driven regime segments of the economy's trajectory must also match asymptotically, as follows from \eqref{eq_res_full_clearing}. Consequently, the economy's average growth rate across supply-driven segments is equal to the average growth rate across demand-driven segments. As the economy expands at $R$ in the supply-driven regime, the same growth rate is achieved, on average, across the demand-driven segments,\footnote{The demand-driven economy cannot reach the asymptotic growth rate $R^\star>R$ as a result of the segments' finite duration and an adverse growth bias due to the typical alignment of demand-driven segments with recessionary periods.} meaning $R$ is also the overall rate of expansion. Figure \ref{fig_results_asymp_general} displays a simulation capturing the realized asymptotic growth path in the general case and highlights the interplay of the supply- and demand-driven dynamics. 

To sum up, the asymptotic growth rates in the demand-driven regime depend on the mechanism underlying economic fluctuations. Fluctuations driven by a limit cycle cannot be realized since they do not satisfy the convergence between supply and demand in the long run. The economy's trajectory realized in the general case of the Dynamic Solow model consists of a chain of supply-driven regimes in which the economy experiences the equilibrium growth and demand-driven regimes in which fluctuations emerge via coherence resonance. Overall, the economy grows asymptotically at the classic Solow rate $R$. Although fluctuations can cause large excursions from this equilibrium growth path, the deviations disappear on the large timescales relevant for the convergence of supply and demand.  

\subsection{Business Cycle Dynamics}\label{sec_results_cycle}

Our analysis of asymptotic growth in the preceding section has led us to conclude that coherence resonance is the relevant endogenous mechanism underlying economic dynamics as it enables the convergence of capital demand and supply over the long run. In this section, we focus on the intermediate timescale to examine endogenous fluctuations produced by the Dynamic Solow model \eqref{eq_res_full_y_dot}-\eqref{eq_res_full_matching} in the coherence resonance regime. 


\begin{figure}[hbt!]
	\centering
	\includegraphics[width=0.8\textwidth]{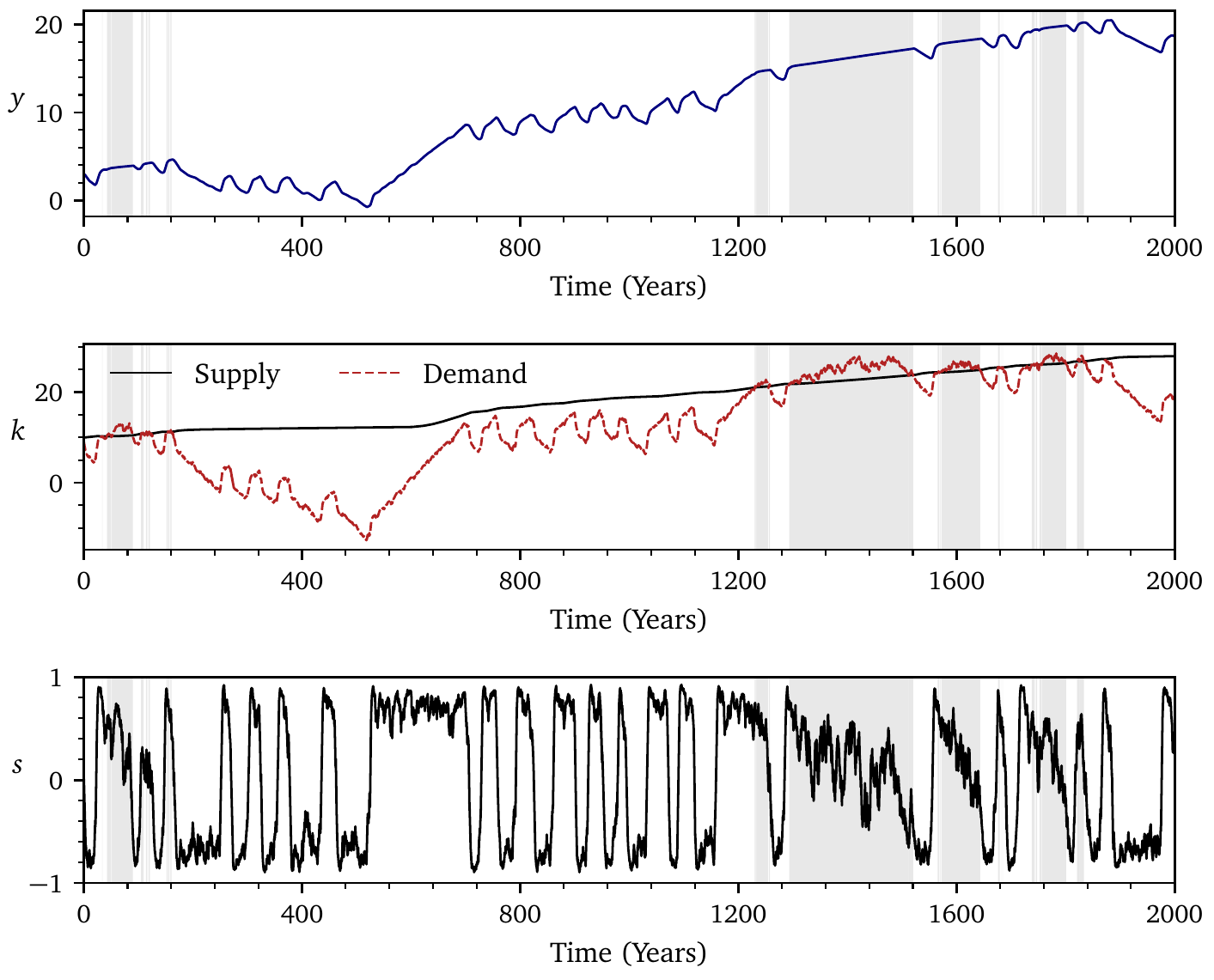}
	\caption{A simulation of the economy's trajectory over the medium term in the model's general case, which highlights a sequence of supply- and demand-driven regimes. Shaded segments correspond to periods during which $k_d>k_s$. Parameters are from the base case in \ref{appx_notation}}.
	\label{fig_results_dynamics}
\end{figure}

Figure \ref{fig_results_dynamics} depicts a typical realization of the economy's trajectory over the medium term. The economy undergoes a sequence of supply- and demand-driven dynamic behaviors, as indicated, respectively, by shaded and unshaded segments. In the demand-driven case, in which demand is below supply, sentiment (lower panel) exhibits distinctively bi-stable behavior, staying for long periods near the positive (expansion) and negative (contraction) equilibria and traversing quickly the distance between them during economic regime transitions. This sentiment behavior leads to fluctuations in demand (middle panel) that, in turn, induce business cycles around the long-term growth trend (upper panel). Conversely, during periods when supply is the limiting factor, sentiment follows a random walk due to the absence of economic feedback and the supply-driven economy exhibits the equilibrium growth dynamics. 

The long-term simulations demonstrate that demand stays below supply on average $\sim70\%$ of the time. This can be interpreted as the firms' decision to hold excess capital \citep[as, for example, noted in][]{Fair2020} as the entire capital supply is made available to firms, implying a capital utilization rate below 100\% over extended periods.\footnote{The utilization rate measures current production as a proportion of potential production. In our context, the analogous measure is the ratio of capital in production in the demand-driven regime $(k=k_d)$ to capital available $(k_s)$. For empirical evidence of firms holding excess capital, see the FRED series ``Capacity Utilization: Total Index'' (CAPUTLB50001SQ) or the Census Bureau's National Emergency Utilization Rate. See also \citet{Murphy2017} for a study of the persistence of excess capital.}

\begin{figure}[htpb!]
    \centering
    \includegraphics[width=1.0\textwidth]{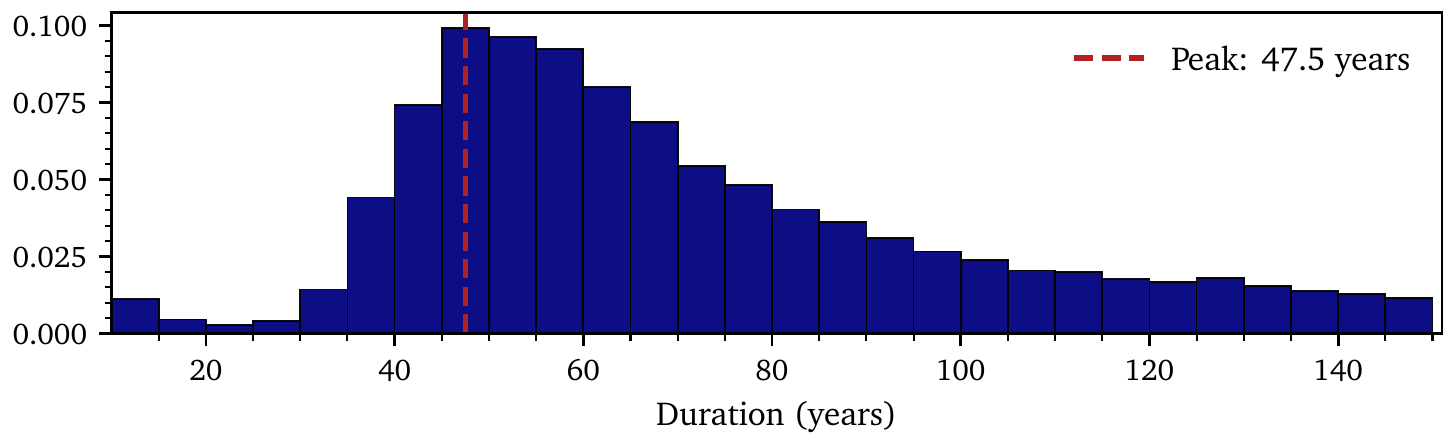}
    \caption{Histogram of the duration of simulated business cycles. The cycles are based on the detrended production, $y-Rt$, with respect to the best fit straight line which coincides with the equilibrium growth path, given by the Solow growth rate $R$ (Section \ref{sec_results_asymp_gc}). Duration is calculated as the time interval between two successive zero crossings in the same direction by detrended production. The histogram, based on the 5 year bins, is truncated at 10 years to eliminate noise artifacts and at 150 years to highlight its peak. The parameters are from the base case in \ref{appx_notation}. 
    }
    \label{fig_results_cycle_duration_production}
\end{figure}

\begin{figure}[htpb!]
    \centering
    \includegraphics[width=1.0\textwidth]{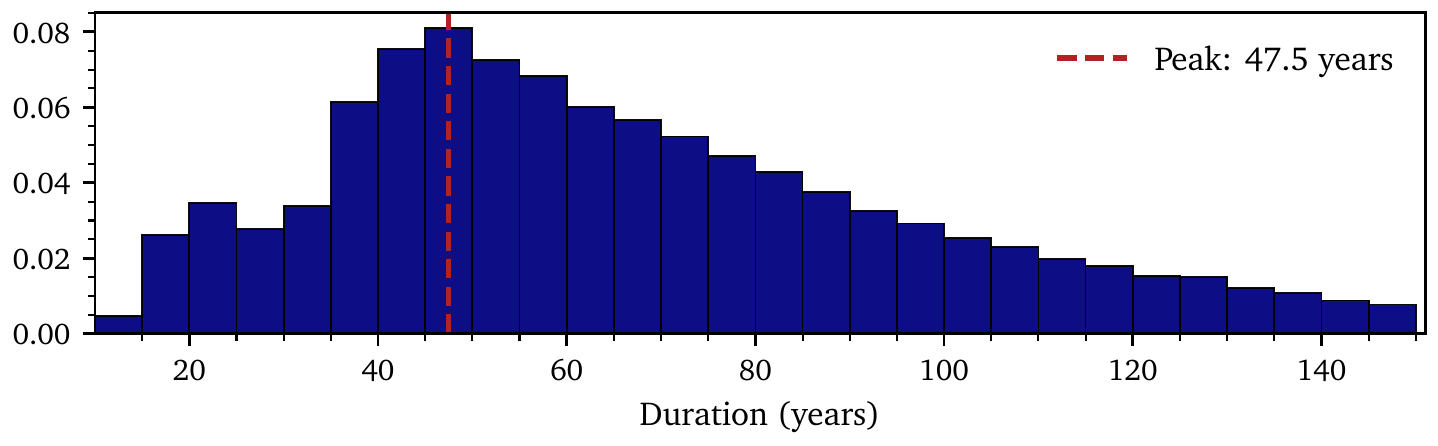}
    \caption{Histogram of the duration of simulated sentiment cycles. The cycles are defined as the roundtrip of sentiment between the positive and negative equilibria. Duration is calculated as the time interval between two successive zero crossings in the same direction by sentiment $s$ in the enforced demand-driven case. The histogram, based on the 5 year bins, is truncated at 10 years to eliminate noise artifacts and at 150 years to highlight its peak. The parameters are from the base case in \ref{appx_notation}. 
    }
    \label{fig_results_cycle_duration_sentiment}
\end{figure}


Figure \ref{fig_results_cycle_duration_production} is a histogram of business cycle periods simulated by the model. It displays a wide distribution with a peak in the 40-70 year interval (with over 50\% of the periods falling into this range), indicating the presence of quasiperiodic fluctuations. To confirm the source of these fluctuations, we inspect the distribution of the lengths of sentiment cycles, defined as the roundtrip of sentiment between the positive and negative equilibria (such as those depicted in the lower panel in Figure \ref{fig_results_dynamics}). This distribution, shown in Figure \ref{fig_results_cycle_duration_sentiment}, also peaks at 40-70 years. It follows that business cycles are, as expected, linked to sentiment transitions from one equilibrium to the other driven by coherence resonance. Therefore, we affirm coherence resonance is the relevant mechanism forming the quasiperiodic fluctuations in output captured in Figure \ref{fig_results_cycle_duration_production}.       

In \ref{appx_parameters}, we show that parameter $c_2$, which defines the sensitivity of capital demand to sentiment, is key to the business cycle duration: the lower $c_2$, the shorter the average duration of business cycles. We also show there that the model admits coherence resonance only if $c_2$ is above a certain critical value and tune the model to be in a regime with $c_2$ close to this value. It follows that coherence resonance -- as a mechanism of business cycles driven by firms' investment -- imposes a natural minimum duration threshold, ruling out fluctuations with a characteristic timespan shorter than the Kondratieff-like 40-70 years.

In current literature, business cycles are typically estimated to last 8-12 years. However, a direct comparison of the duration would be misleading as our model, centered on capital demand dynamics, does not include links to the faster-paced processes, such as credit or equity market dynamics, that can accelerate business cycles through further interactions with the real economy. In other words, our model captures capital demand driven cycles, which are arguably just one of a number of fluctuation modes that reinforce or otherwise affect each other to produce the business cycles observed in the real world.

On that point, we take note of \citet{KroujilineEtAl2019} that studies combined effects in a coupled macroeconomic system, attaching the interactions-based stock market model of \citet{GusevEtAl2015} (capable of producing relatively short-term endogenous cycles) to the simple phenomenological model of the economy of \citet{Blanchard1981} (within which output follows slow relaxation dynamics) to obtain quasiperiodic fluctuations with the same frequency as observed business cycles. A natural next step would be to investigate whether a more advanced coupled system, where both the financial sector and the real economy experience nonlinear endogenous dynamics at different frequencies
can replicate and explain observed macroeconomic behaviors in greater detail.\footnote{Such as the stock market model of \citet{GusevEtAl2015} and the present Dynamic Solow model of the economy, which share a similar framework for micro-level interactions.}


\section{Conclusion}\label{sec_conclusion}

In this paper we have developed the Dynamic Solow model, a tractable macroeconomic model that captures dynamic behaviors across multiple timescales, and applied it to study economic fluctuations and their impact on long-term growth. 

Our model consists of a dynamic capital demand component, representing an interactions-based process whereby firms determine capital needs and make investment decisions, and a simple capital supply component in the form of a Solow growth economy. These components are interlinked via a capital market, which comprises a local inelastic market clearing condition reflecting short-term price rigidity and an asymptotic market clearing condition valid for timescales in which prices are sufficiently flexible to match supply and demand. Starting from the micro-level interactions among firms, we derived the macroscopic equations for capital demand that constitute the dynamic core of the model and attached them to a capital motion equation for a static representative household (providing a dynamic version of the Cobb-Douglas production equation to capture short-term processes) and the aforementioned market clearing conditions. As a result, we have obtained a closed-form nonlinear dynamical system that allows for the examination of a broad range of economic behaviors. 

The Dynamic Solow model admits two characteristic regimes, depending on whether capital demand exceeds supply or not. When demand exceeds supply, supply drives output and the dynamic demand component decouples from the rest of the economy, placing the economy on the familiar equilibrium growth path. Otherwise, demand drives output and the model is shown, for economically realistic parameters, to possess two attracting equilibria, one where the economy contracts and the other where it expands. This bi-stable geometry gives rise to business cycles manifested as endogenous fluctuations, wherein the economy’s long entrapment in recessions and expansions is punctuated by rapid alternations between them. We show that, in our model, the economy’s realized trajectory is forged by an interplay of these regimes such that the supply-driven equilibrium dynamics and demand-driven fluctuations continuously succeed one another. We further show that the economy spends around 70\% of its time in the demand-driven regime, indicating fluctuations represent a prevalent economic behavior.

We identify a coherence resonance phenomenon, whereby noise applied to a dynamical system leads to a quasiperiodic response, to be the mechanism behind demand-driven fluctuations. In our model, exogenous noise (representing news received by analysts) instigates the economy’s transition from one equilibrium to the other, resulting in recurrent booms and busts. As such, news shocks act as a catalyst, which is compatible with the ``small shocks, large cycle'' effect observed in the real-world economy. In addition, under a different range of parameter values, we obtain a stochastic limit cycle (i.e. a limit cycle perturbed by exogenous noise) likewise capable of generating endogenous fluctuations. We show, however, that this type of fluctuations cannot be realized as the growth dynamics induced by it do not allow supply and demand to converge in the long run. While both limit cycle and coherence resonance mechanisms are hardwired in our model, in the sense that the parameter ranges must be appropriately selected, we conjecture that in reality the economy self-regulates towards the coherence resonance parameter ranges via long-term price adjustment responsible for the convergence of supply and demand in the long run.

The distribution of the business cycle periods simulated by our model displays a peak in the Kondratieff range of 40-70 years, demonstrating the quasiperiodic character of demand-driven fluctuations. We further find coherence resonance imposes a minimum duration threshold that rules out fluctuations peaking at shorter lengths. This result seems sensible because our model, centered on capital demand dynamics, has no links to faster-paced processes (such as credit or equity market dynamics) that can accelerate fluctuations to be in line with the observed business cycles. A natural extension would be to develop a coupled system, within which both the financial sector representing such faster-paced processes and the real economy experience nonlinear endogenous dynamics at different characteristic frequencies.

Our simulations show that although demand-driven fluctuations occasionally cause large excursions from the equilibrium growth path, the deviations vanish in the long run as supply and demand converge. In our model, the equilibrium growth path is defined by the Solow growth rate in which technology growth appears, simplistically, as a fixed exogenous parameter. From this perspective, endogenizing technological progress in the model may help better understand the long-term impact of fluctuations on growth, presenting an intriguing topic for future research.

\section{Acknowledgments}
\label{sec_ack}
We deeply thank J.-P. Bouchaud who contributed to the early stages of this work and Dmitry Ushanov for contributing ideas on the numerical analysis of the model. We also thank Mikhail Romanov for helpful comments and suggestions. This research was conducted within LGT Capital Partners and the Econophysics \& Complex Systems Research Chair, the latter under the aegis of the Fondation du Risque, the Fondation de l’Ecole polytechnique, the Ecole polytechnique and Capital Fund Management. Maxim Gusev and Dimitri Kroujiline are grateful to LGT Capital Partners for financial support. Karl Naumann-Woleske also acknowledges the support from the New Approaches to Economic Challenges Unit at the Organization for Economic Cooperation and Development (OECD).

\newpage
\bibliographystyle{elsarticle-harv}
\bibliography{References.bib}

\appendix
\section{Model Variables and Parameters}\label{appx_notation}
\begin{table}[H]
\centering
\caption{Table of Notation} \label{tab_notation}
\begin{tabular}{lrrr}
\toprule
{} & {Definition} & {Type} & {Base Case Value}\\
\midrule
\multicolumn{4}{l}{\textbf{Production (Section \ref{sec_model_production})}}\\
$Y$    &  Production  & Variable &- \\
$K$    &  Capital in production & Variable & - \\
$y$    &  Log production  & Variable &- \\
$k$    &  Log capital in production & Variable & - \\
$\tau_Y$    & Production characteristic timescale (business days) & Constant & $1000$ \\
$A_0$    & Initial technology level  & Constant &  1 \\
$\varepsilon$    & Daily technology growth rate  & Constant & $2.5\times10^{-5}$ \\
$\rho$    &  Capital share in production & Constant  &  $1/3$ \\\\
\multicolumn{4}{l}{\textbf{Capital Supply (Section \ref{sec_model_household})}}\\
$K_s$    &  Capital supply  & Variable &- \\
$k_s$    &  Log capital supply  & Variable &- \\
$\lambda$    & Proportion of income saved  & Constant &  $0.15$ \\
$\delta$    & Daily depreciation rate  & Constant &  $2\times10^{-4}$ \\\\
\multicolumn{4}{l}{\textbf{Capital Demand (Section \ref{sec_model_demand})}}\\
$K_d$    &  Capital demand  & Variable &- \\
$k_d$    &  Log capital demand  & Variable &- \\
$s$    &  Sentiment level  & Variable &- \\
$h$    &  Information level  & Variable &- \\
$c_1$    &  Capital demand sensitivity to $\dot{s}$ & Constant & $3$ \\
$c_2$    &  Capital demand sensitivity to $s$ & Constant & $7\times10^{-4}$ \\
$\beta_1$    &  Sentiment herding factor  &  Constant& $1.1$ \\
$\beta_2$    &  Sentiment sensitivity to $h$ &  Constant& $1.0$ \\
$\gamma$    &  Feedback strength factor  &  Constant& $2000$ \\
$\xi_t$    &  News shocks  & Exogenous noise &- \\
$\tau_s$    & Sentiment characteristic timescale (business days) & Constant & $250$ \\
$\tau_h$    & Information characteristic timescale (business days)  & Constant & $25$ \\
$\tau_\xi$    & News shocks characteristic timescale (business days) & Constant & $5$ \\\\
\multicolumn{4}{l}{\textbf{Limiting Cases (Section \ref{sec_limitcases})}}\\
$z$    & Production growth indicator & Variable &- \\
$\omega_y$ & $\omega_y=\tau_y^{-1}$ & Constant & 0.001\\\\

\multicolumn{4}{l}{\textbf{General Case (Section \ref{sec_results})}}\\
$y_0, k_{s0}, k_{d0}$ & Asymptotic growth rates of log variables & Variable & -\\
$R$ & Classic Solow growth rate $R=\varepsilon/(1-\rho)$ (daily) & Constant & $3.75\times10^{-5}$\\
$1/\varepsilon$ &Technology growth timescale (business days) & Constant & $4\times10^{4}$\\
\bottomrule
\end{tabular}
\end{table}

\section{Approximate Solution in the Supply-Driven Regime}\label{appx_boundarylayer}
In this appendix, we solve equation \eqref{eq_limit_ks_second_order} approximately through use of the boundary layer technique and obtain the economy's path in analytic form in the intermediate and long run under the supply-driven regime ($K=K_s$). 

The starting point of our derivation is equation \eqref{eq_limit_ks_second_order}, for convenience repeated here:
\begin{equation}\label{eq_appx_limit_ks_second_order}
    \tau_Y\ddot{K} + (1+\tau_Y\delta)\dot{K} + \delta K = \lambda K^\rho e^{\varepsilon t}.
\end{equation}
Recall that $1\ll\tau_Y\ll1/\varepsilon$, where $\tau_Y$ is the timescale in which output adjusts to changes in the level of capital and $1/\varepsilon$ is the timescale of output growth in the long run. We aim to capture the dynamics on these two timescales by solving equation \eqref{eq_appx_limit_ks_second_order} on the interval $t\geq O(\tau_y)$. For simplicity, we assume that $\tau_Y\delta\gg1$, which implies that $\tau_Y\delta\dot{K}$ is much larger than $\dot{K}$ and $\tau_Y\ddot{K}$ on the interval $t\geq O(\tau_y)$, allowing us to derive a more compact solution.

First, we consider equation \eqref{eq_appx_limit_ks_second_order} for $t\gg\tau_y$. In this outer region, $\tau_y\dot{K}\ll K$ and we can approximate the solution to \eqref{eq_appx_limit_ks_second_order} by the solution to equation:
\begin{equation}
\delta K_o = \lambda K_o^\rho e^{\varepsilon t},
\end{equation}
which is given by
\begin{equation}\label{eq_appx_outer}
    K_o = \left(\frac{\lambda}{\delta}e^{\varepsilon t}\right)^{\frac{1}{1-\rho}}.
\end{equation}

Next, we consider equation \eqref{eq_appx_limit_ks_second_order} on the interval $O(\tau_y)\leq t\ll1/\varepsilon$, where $e^{\varepsilon t}\rightarrow1$ and $\tau_y\dot{K}$ is not necessarily substantively smaller than $K$. In this inner region, we can approximate the solution to \eqref{eq_appx_limit_ks_second_order} by the solution to
\begin{equation}
    \tau_y\dot{K}_i + K_i = \frac{\lambda}{\delta}K^\rho_i.
\end{equation}
This is the Bernoulli equation and its solution is given by 
\begin{equation}\label{eq_appx_inner}
    K_i = \left(B e^{-(1-\rho)\frac{t}{\tau_y}}+\frac{\lambda}{\delta}\right)^{\frac{1}{1-\rho}},
\end{equation}
where $B$ is the constant of integration. 

Solutions $K_o$ and $K_i$ must match in the overlapping interval $\tau_Y\ll t\ll1/\varepsilon$. This is satisfied for any value of $B$ since $K_o\rightarrow\left(\frac{\lambda}{\delta}\right)^{\frac{1}{1-\rho}}$ 
and $K_i\rightarrow\left(\frac{\lambda}{\delta}\right)^{\frac{1}{1-\rho}}$, as follows from \eqref{eq_appx_outer} and \eqref{eq_appx_inner}. Thus, we approximate the solution to equation \eqref{eq_appx_limit_ks_second_order} in this region by
\begin{equation}
K_m = \left(\frac{\lambda}{\delta}\right)^{\frac{1}{1-\rho}}.
\end{equation}
The approximate solution to equation \eqref{eq_appx_limit_ks_second_order} that is uniformly valid for all $t\geq O(\tau_y)$ is given by
\begin{equation}\label{eq_appx_cap_sol}
    K = K_i+K_o-K_m = \left(\frac{\lambda}{\delta}\right)^{\frac{1}{1-\rho}}\left(\left(Be^{-\left(\frac{1-\rho}{\tau_y}\right)t}+1\right)^{\frac{1}{1-\rho}}+e^{\left(\frac{\varepsilon}{1-\rho}\right)t}-1\right),
\end{equation}
where $B$ has been rescaled for convenience.

As a final step, we obtain the solution for output $Y$ by inverting the equation of capital motion \eqref{eq_cap_sup}:
\begin{equation}
	Y = \frac{1}{\lambda}\left(\dot{K}+\delta K\right).
\end{equation}

Note that $\dot{K}\ll\delta K$ on the interval $t\geq O(\tau_Y)$ due to the simplifying assumption $\tau_Y\delta\gg1$. Therefore, the corresponding uniform approximation for output $Y$, valid for all $t\geq O(\tau_Y)$, is given by

\begin{equation}\label{eq_appx_prod_sol}
    Y = \left(\frac{\lambda}{\delta}\right)^{\frac{\rho}{1-\rho}}\left(\left(Be^{-\left(\frac{1-\rho}{\tau_y}\right)t}+1\right)^{\frac{1}{1-\rho}}+e^{\left(\frac{\varepsilon}{1-\rho}\right)t}-1\right).
\end{equation}

\section{Model Parameterization}\label{appx_parameters}
In this appendix, we examine the model's parameters and discuss how they affect the behavior of the dynamical system \eqref{eq_kd_shz_system} in the phase space.

We begin with equation \eqref{eq_z_s_dot} that describes sentiment dynamics. Parameter $\beta_1$ defines the relative importance of the herding and random behaviors of firms. In an unforced situation ($\beta_2=0$), the number of stable equilibrium points, to which the firms' sentiment $s$ converges, doubles at $\beta_1=1$ from one to two. For $\beta_1<1$, random behavior prevails since there is a single equilibrium at $s=0$, meaning firms fail to reach a consensus opinion. Conversely, for $\beta_1>1$, herding behavior rules as equation \eqref{eq_z_s_dot} generates a polarized, bi-stable environment with one pessimistic ($s<0$) and one optimistic ($s>0$) equilibrium states. It is sensible to assume $\beta_1\sim1$, otherwise firms would unrealistically behave either randomly or in perfect synchronicity. We set $\beta_1=1.1$, implying a slight prevalence of herding over randomness. In addition, we set $\beta_2=1$ to ensure that analysts' influence on firms' managers likewise appears in the leading order. 

We now consider the information dynamics in \eqref{eq_z_h_dot}. The terms under the hyperbolic tangent describe the impacts of economic growth and exogenous news on the collective opinion of analysts $h$. We assume these two sources of information are of equal importance. Thus, we expect that $\gamma\omega_y=O(1)$ in the feedback term and we model $\xi_t$ as an Ornstein-Uhlenbeck process with an $O(1)$ standard deviation and short decorrelation timescale $\tau_\xi$. Note that $\omega_y\ll1$ and accordingly $\gamma\gg1$. 

Finally, we inspect the economic dynamics in \eqref{eq_z_z_dot}. In this equation, different terms determine leading behaviors on separate timescales. We show in \ref{appx_asymp_derivation} that the last three terms (with technology growth rate $\varepsilon$ estimated on the basis of observed total factor productivity) are in balance in the long run. However, if we consider short timescales, the change in sentiment $\dot{s}$ becomes dominant. Thus, equation \eqref{eq_z_z_dot} can be approximated in the short run as $\dot{z}\sim \rho c_1\dot{s}$ and we set $\rho c_1=1$. We also note that by construction $c_2\ll c_1$ to ensure that the term $c_2s$ does not contribute to capital demand dynamics on short timescales. Hence we expect $c_2\ll 1$.  

As highlighted in Section \ref{sec_model}, there is a segregation of characteristic timescales that emerges naturally from the types of decisions faced by the different agents in the model: $\tau_\xi\ll\tau_h\ll\tau_s\ll\tau_y\ll1/\varepsilon$. This segregation facilitates the transfer of the impact of instantaneous news shocks $\xi_t$ across multiple timescales. The estimates for the timescales are discussed in Section \ref{sec_model_capmkt}.

\begin{figure}
	\centering
	\includegraphics[width=\textwidth]{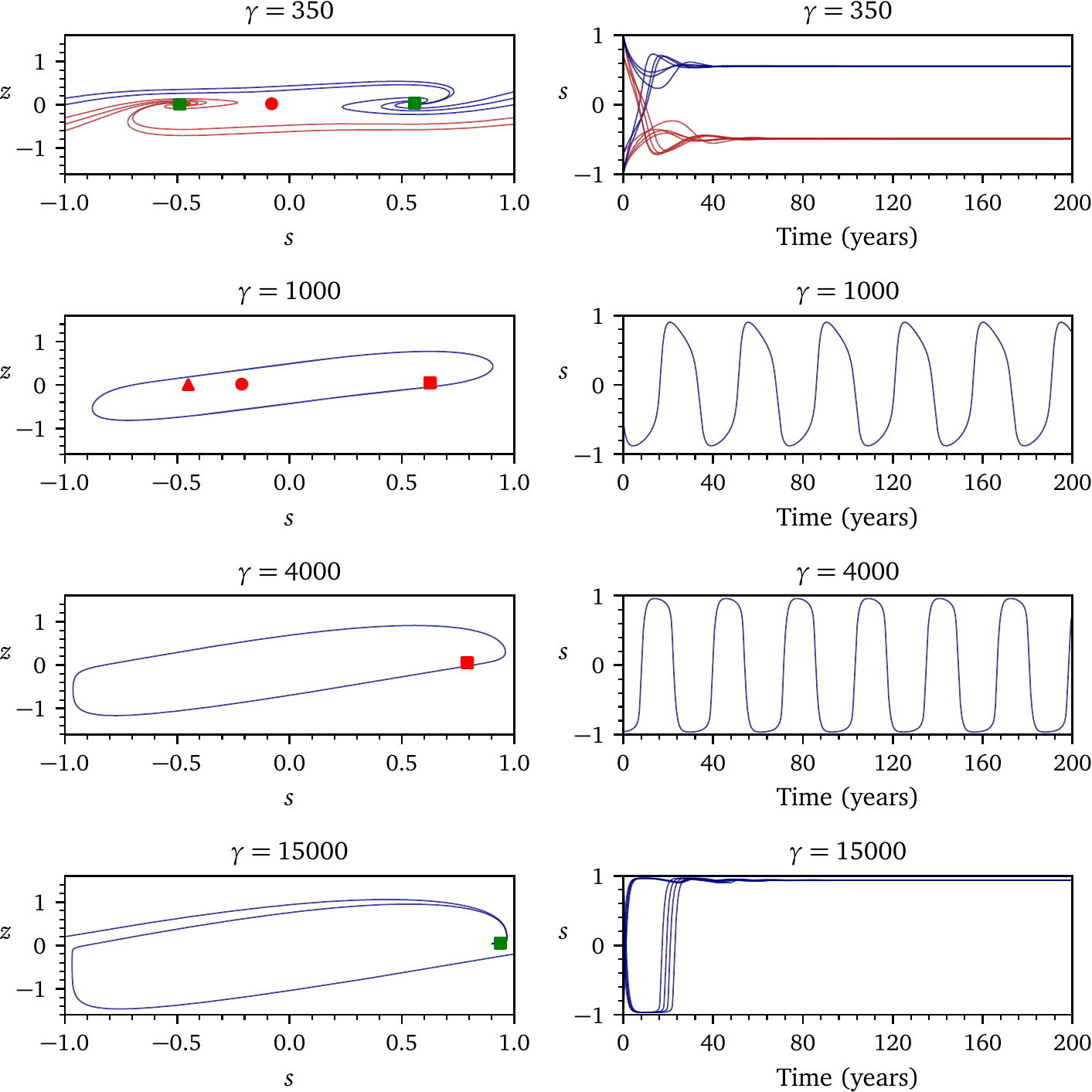}
	\label{fig_appxC_cycle_g}
	\caption{Development of a stable limit cycle with increasing $\gamma$ for $\xi_t=0$, $c_2=1\times10^{-4}$ and all other parameter values from the base case (\ref{appx_notation}). The left panels show the phase portraits projected on the ($s,z$)-plane and the right panels plot $s(t)$. Classification of equilibrium points is provided in footnote \ref{point notatrion} in Section \ref{sec_kd_limit}. As $\gamma$ increases, a large stable limit cycle emerges and then vanishes, demonstrating the destabilizing effect of $\gamma$ at low values and its stabilizing effect at high values. (i) Stable dynamics for $\gamma=350$: red trajectories converge to the left focus and blue trajectories converge to the right focus. (ii) The equilibria become unstable and a large stable limit cycle emerges for $\gamma=1000$. (iii) The left node and the saddle point vanish while the limit cycle persists at $\gamma=4000$. (iv) The dynamics are again stable at $\gamma=15000$: trajectories converge to the stable focus and the limit cycle disappears.} 
	\label{fig_appxC_cycle}
\end{figure}

The parameters $c_2$ and $\gamma$ are central to the system's behavior in the phase space. Increasing $c_2$ stabilizes the system, strengthening convergence towards the stable equilibria and creating a higher barrier between attracting regions. The role of $\gamma$ is twofold. As $\gamma$ grows from zero, its immediate effect is to destabilize the system due to growing economic feedback. However, as $\gamma$ continues to increase, it exerts a stabilizing effect similar to that of $c_2$ because of the term $\gamma c_2$ in the equilibrium condition: 
\begin{equation}\label{eq_kd_equilibrium_condition}
\textrm{arctanh}(s)-\beta_1s = \beta_2\tanh\left(\gamma c_2s + \gamma\varepsilon\right), 
\end{equation}
which follows from equations \eqref{eq_kd_shz_system} for $\dot{h}=\dot{s}=\dot{z}=\xi_t=0$. Consequently, the potential to generate autonomous economic instability is limited. In particular, there exists a critical value\footnote{
	Subject to the values set for the other parameters.
} of $c_2\sim10^{-4}$ below which feedback may generate a limit cycle and above which it does not. Figure \ref{fig_appxC_cycle} depicts the formation and subsequent destruction, for $c_2=10^{-4}$, of the limit cycle as $\gamma$ increases.

\begin{figure}
    \centering
    \includegraphics{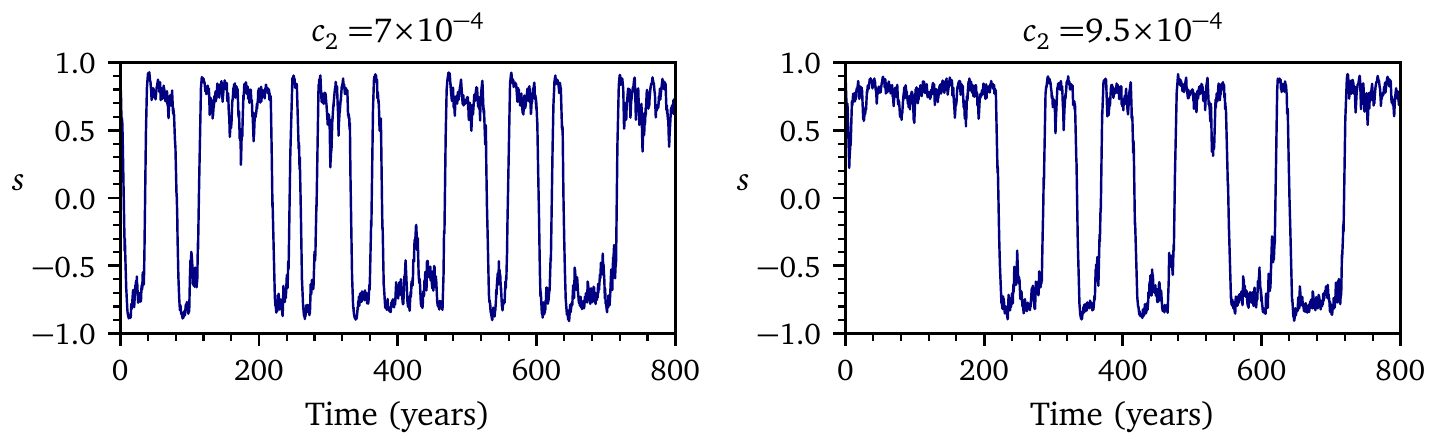}
    \caption{The effect of $c_2$ on the dynamics of sentiment $s(t)$ for $\xi_t\neq0$. As $c_2$ increases from the base case value $c_2=7\times10^{-4}$ (left) to $c_2=9.5\times10^{-4}$ (right), the barrier separating the two attracting regions grows stronger. The system spends more time captive to the attractors, reducing the frequency of the crossings between them and lengthening the duration of fluctuations. Note that the system tends to stay longer at the expansion attractor (where $s>0$) owing to the asymmetry induced by technological growth $\varepsilon>0$. All other parameters are from the base case (\ref{appx_notation}).}
    \label{fig_appxC_base_c2_dynamics}
\end{figure}

In this paper, we argue that realistic economic behaviors cannot be explained by a stochastic limit cycle. Therefore, we proceed to study the system for $c_2\gtrsim10^{-4}$, which ensures a bi-stable configuration without a limit cycle. Figure \ref{fig_appxC_base_c2_dynamics} illustrates that as $c_2$ increases, the barrier between attracting regions grows stronger, resulting in less frequent crossings from one region to the other (i.e. cycle duration increases). We seek $c_2$ at the lower end of this range to reduce cycle duration. 

\begin{figure}
    \centering
    \includegraphics{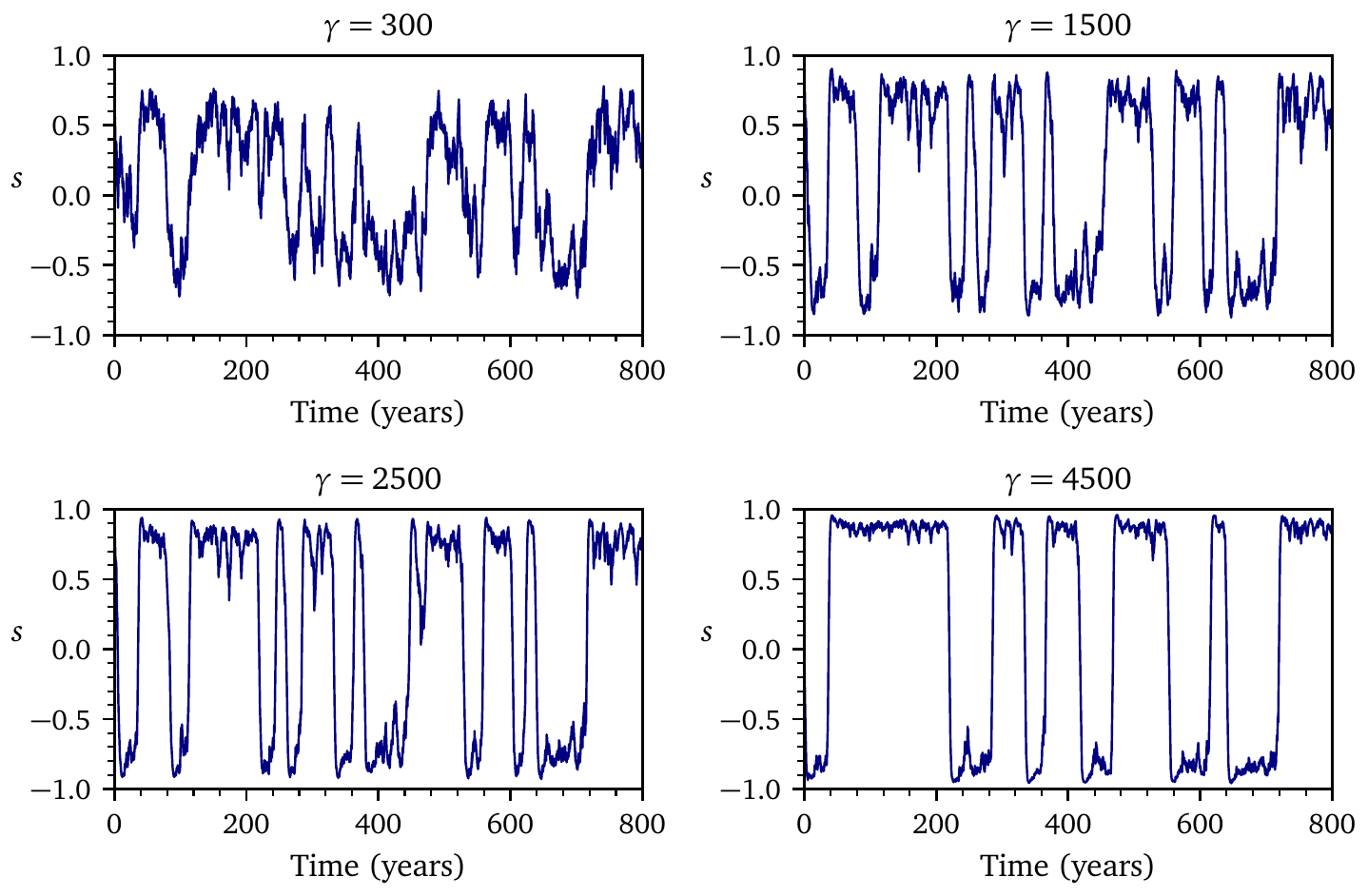}
    \caption{The effect of $\gamma$ on the dynamics of sentiment $s(t)$ for $\xi_t\neq0$. At low $\gamma$, the system's behavior is dominated by noise as the barrier between the two attracting regions is weak. As $\gamma$ increases, the barrier grows stronger and the system becomes extremely bi-stable. Reasonably balanced dynamics emerge in the range from $\gamma=1500$ to $\gamma=2500$. Note the asymmetry caused by technological growth becomes exacerbated as $\gamma$ increases in accordance with equation \eqref{eq_kd_equilibrium_condition}. All other parameters are from the base case (\ref{appx_notation}).}
    \label{fig_appxC_base_gamma_dynamics}
\end{figure}

Similarly, the barrier between attracting regions grows stronger as $\gamma$ increases, resulting in infrequent transitions between the attractors. Relatively small values of $\gamma$, however, dampen feedback, leading to weak dynamics and stochastic-like behavior. Accordingly, we focus on values of $\gamma$ between these two extremes. Figure \ref{fig_appxC_base_gamma_dynamics} depicts the dynamics under different values of $\gamma$, with balanced dynamic behaviors for a reasonably wide range thereof.

\begin{figure}
    \centering
    \includegraphics[width=\textwidth]{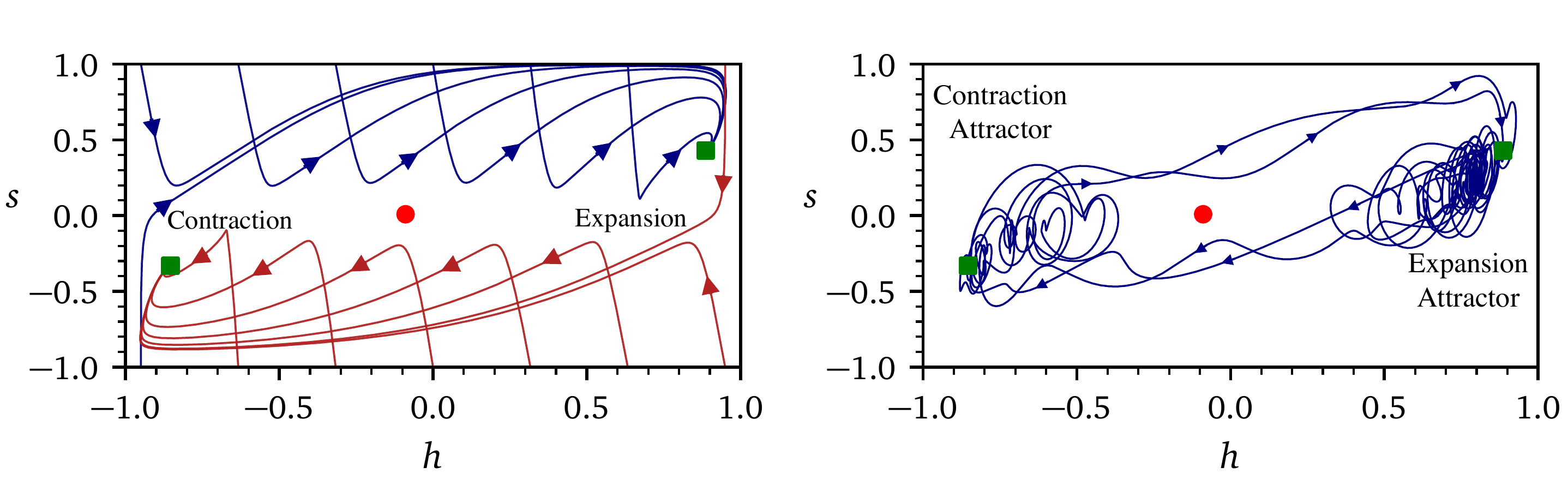}
    \caption{System dynamics in the base case (\ref{appx_notation}). Left: A phase portrait ($\xi_t=0$) projected on the ($s,h$)-plane. The portrait depicts stable foci, separated by a saddle point, and the large trajectories relevant for regime transitions. Right: A trajectory ($\xi_t\neq0$) projected on the ($s,h$)-plane. The stable foci are at the center of the two attracting regions, within which the trajectory is dense. The transit of the economy between these regions corresponds to regime transitions between contractions and expansions, occurring at much shorter intervals than the periods during which the economy is captive to an attractor. The trajectory was smoothed by a Fourier filter to remove harmonics with periods less than 500 business days for clean visualization.}
    \label{fig_appxC_base_sh_phase}
\end{figure}

We select $c_2=7\times10^{-4}$ and $\gamma=2000$ for the base case studied in Sections \ref{sec_kd_limit} and \ref{sec_results}. Note that $c_2\ll 1$, $\gamma\gg1$ and $\gamma\omega_y=O(1)$, as expected. Figure \ref{fig_appxC_base_sh_phase} provides the base case phase portrait ($\xi_t=0$) projected on the $(s,h)-$plane, showing attracting regions around the two stable equilibria as well as long trajectories passing near each attractor and ending at the opposite equilibrium. These trajectories, which emerge due to strong feedback ($\gamma\gg1$), allow the economy to transition quickly between expansions and contractions. 

\begin{figure}
    \centering
    \includegraphics{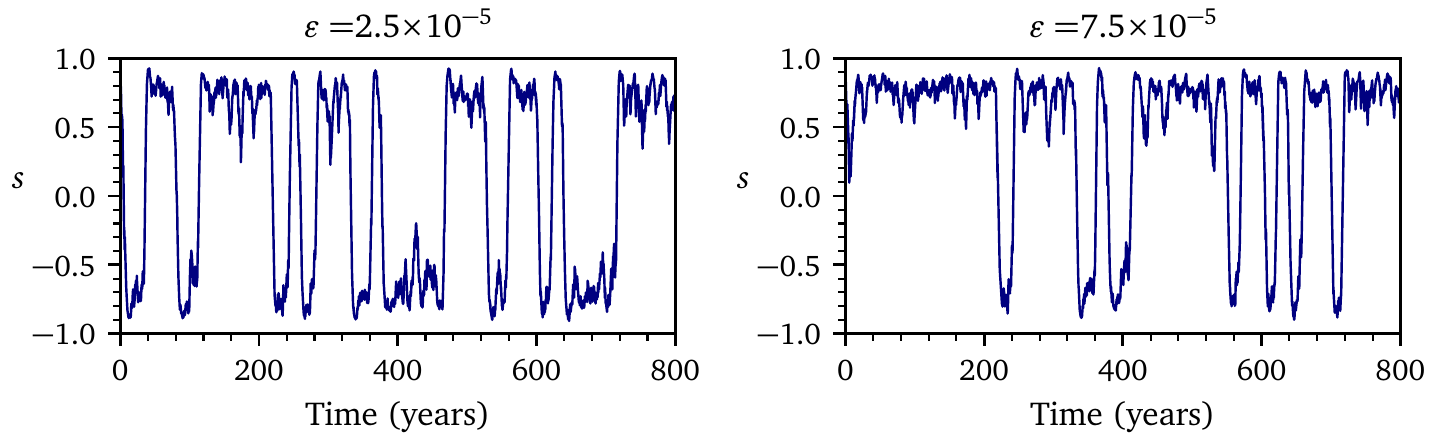}
    \caption{The effect of $\varepsilon$ on the dynamics of sentiment $s(t)$ for $\xi_t\neq0$. As $\varepsilon$ increases from the base case value $\varepsilon=2.5\times10^{-5}$ (left) to $\varepsilon=7.5\times10^{-5}$ (right), the system behavior begins to exhibit a stronger asymmetry between the contraction and expansion attractors. All other parameters are from the base case (\ref{appx_notation}).}
    \label{fig_appxC_base_epsilon_asymmetry}
\end{figure}

The attractors are not connected and the economy cannot cross the boundary separating them in the absence of exogenous news shocks $\xi_t$. It takes a random news event to force the economy, entrapped by one attractor, across the boundary. Once it crosses the boundary, the economy finds itself on the long trajectory that takes it swiftly to the other attractor, where the economy remains captive until another news event instigates the next regime transition by again forcing the economy across the boundary. At this point the economy is carried back to the entrapment region where it started. This is the coherence resonance mechanism that is at the heart of the economic fluctuations in our model. 

As a final comment, we note that if $\varepsilon=0$, the equilibrium condition \eqref{eq_kd_equilibrium_condition} is symmetric to $s\to -s$. Technology growth, $\varepsilon>0$, causes an asymmetry\footnote{Note that $\gamma$ amplifies the asymmetry via the term $\gamma\varepsilon$ in equation \eqref{eq_kd_equilibrium_condition}.} wherein the equilibrium at $s>0$ becomes stronger than that at $s<0$ (to the extent that the latter vanishes above a certain threshold). As a result, the system tends to stay longer in the region where economic sentiment is positive, accelerating the economy's long-term growth. The asymmetry, however, vanishes in the limit cycle regime, whether periodic or stochastic (Section \ref{sec_results_asymp}). Figure \ref{fig_appxC_base_epsilon_asymmetry} illustrates this asymmetric behavior.

\section{Asymptotic Analysis of Long-Term Growth}\label{appx_asymp_derivation}

In this appendix, we study the behavior of the Dynamic Solow model in the long run by seeking $y \sim y_0 t$, $k_d \sim k_{d0} t$ and $k_s \sim k_{s0} t$ in equations \eqref{eq_res_full_y_dot}-\eqref{eq_res_full_matching} at large values of $t$.

\subsection{Asymptotic Behavior in the Supply-Driven Regime ($k_d>k_s$)}\label{subsec_supply}
We first consider the situation where capital demand exceeds supply, which entails $k=k_{s}$ under the market clearing condition \eqref{eq_res_full_clearing}, and obtain the resulting growth rates. 

For $t\gg1$, the production equation \eqref{eq_res_full_y_dot} becomes
\begin{equation}
	e^{(\rho k_{s0}+\varepsilon - y_0) t} - 1 = \tau_y y_0.
\end{equation}
Consequently, $(\rho k_{s0} + \varepsilon - y_0 )t$ must be constant, which in turn implies that 
\begin{equation}\label{eq_lt_supply_y0}
	y_0 = \rho k_{s0} + \varepsilon,
\end{equation}
with a precision of up to $O(1/t)$. Similarly, capital supply equation \eqref{eq_res_full_ks_dot} yields
\begin{equation}
	k_{s0} = \lambda e^{(y_0-k_{s0})t} - \delta,
\end{equation}
so that $(y_0 - k_{s0})t$ is constant and, therefore, with a precision of up to $O(1/t)$:
\begin{equation}\label{eq_lt_supply_ks0}
    k_{s0} = y_0.
\end{equation}
It follows from equations \ref{eq_lt_supply_y0} and \ref{eq_lt_supply_ks0} that
\begin{equation}\label{y0_solow}
	y_0 = k_{s0} = \frac{\varepsilon}{1-\rho} \equiv R,
\end{equation}
where $R$ denotes the classic Solow growth rate.\footnote{This same result also follows from equation \eqref{eq_bla} for $t{\geq}O(1/\varepsilon)$.} 

To determine the growth rate of capital demand $k_{d0}$, we average equation \eqref{eq_res_full_kd_dot} with respect to time, noting that $\bar{\dot{s}}=0$ since $s$ is bounded:
\begin{equation}\label{eq_lt_supply_kd0}
	k_{d0} = c_2 \bar{s},
\end{equation}
where the bar denotes the time average.

Then we average equation \eqref{eq_res_full_h_dot} while noting that $\bar{\dot{h}}=0$ since $h$ is bounded and that $H(k_s,k_d)=0$ from \eqref{eq_H_switch} (no feedback) to obtain
\begin{equation}\label{eq_lt_supply_h}
	\bar{h} = \overline{\tanh\left(\xi_t\right)} = \tanh\left(\overline{\xi_t}\right) = 0,
\end{equation}
where we have assumed that fluctuations are small to allow us to take averages under the hyperbolic tangent\footnote{This simplifying assumption does not severely restrict applicability as $\tanh{x}$ is approximated reasonably well by a linear function for $-1{\leq}x{\leq}1$ and the noise amplitude is $O(1)$ in \eqref{eq_lt_supply_h}.}. Similarly averaging equation \eqref{eq_res_full_s_dot} leads to
\begin{equation}\label{eq_lt_supply_s}
	\bar{s} = \overline{\tanh\left(\beta_1 s + \beta_2 h\right)} = \tanh\left(\beta_1 \bar{s}\right).
\end{equation}

Equation \eqref{eq_lt_supply_s} has three solutions for $\beta_1>1$: $s=0$, $s_{-}<0$, and $s_{+}>0$, where $s_{-}=-s_{+}$. Our focus is on $s_{-}$ and $s_{+}$ as they correspond to the stable equilibrium points\footnote{For the base case value $\beta_1=1.1$, we have $s_{\pm}\approx\pm 0.5$ from \eqref{eq_lt_supply_s}.}. The system spends most of its time in the attracting regions that surround each of these two equilibria and transits rapidly between them when forced by exogenous noise. In the long run, the time spent in transit is negligible relative to the length of time during which the system is entrapped by the attractors. The attractors have the same strength and are located symmetrically in $s$, thus the system tends to spend an equal amount of time at each of them at large $t$. Therefore, its average position with respect to sentiment $s$ must be zero. More formally, taking $s_{-}$ and $s_{+}$ as the attractors' proxies, we estimate the long-term average sentiment as
\begin{eqnarray}\label{sbar_sup_est}
	\bar{s} = \frac{1}{2}\left(s_{-}+s_{+}\right) = 0. 
\end{eqnarray}
Hence equation \eqref{eq_lt_supply_kd0} yields
\begin{equation}
	k_{d0} = 0.
\end{equation}

This result is intuitively clear: the growth of demand is driven in the long run by average sentiment, which converges to zero because its dynamics are symmetric in the absence of feedback. We conclude that in the supply-driven regime the economy's growth is, as expected, independent of capital demand and matches the classic Solow growth, $y_0=k_{s0}=R$, while capital demand is stagnating ($k_{d0}=0$). We verify these results via numerical simulations in Section \ref{sec_results_asymp}.


\subsection{Asymptotic Behavior in the Demand-Driven Regime ($k_d<k_s$)}
In the demand-driven regime, the market clearing condition \eqref{eq_res_full_clearing} yields $k=k_{d}$, so that equation \eqref{eq_res_full_y_dot} becomes
\begin{equation}
e^{(\rho k_{d0}+\varepsilon - y_0) t} - 1 = \tau_y y_0.
\end{equation}
Consequently, 
\begin{equation}\label{eq_lt_demand_y0}
y_0 = \rho k_{d0}+\varepsilon,
\end{equation}
with a precision of up to $O(1/t)$. Similarly, equation \eqref{eq_res_full_ks_dot} takes the form: 
\begin{equation}
k_{s0} = \lambda e^{(y_0-k_{s0})t} - \delta e^{k_d-k_s}.
\end{equation}
The term $\delta e^{k_d-k_s}$ can be neglected as it is exponentially small for $k_d<k_s$; therefore, with a precision of up to $O(1/t)$:
\begin{equation}\label{eq_lt_demand_ks0}
y_0 = k_{s0}.
\end{equation}
And finally, averaging equation \eqref{eq_res_full_kd_dot} leads to
\begin{equation}\label{eq_lt_demand_kd0}
	k_{d0} = c_2 \bar{s}.
\end{equation}

We can rewrite equations \eqref{eq_lt_demand_y0} and \eqref{eq_lt_demand_ks0} as
\begin{equation}
y_0 = k_{s0} = R + \rho\left(k_{d0} - R\right).
\end{equation}
It follows that if $\bar{s}>\frac{R}{c_2}$, then the economy's long-term growth exceeds the classic Solow growth rate $R$. For the base case values of $c_2$, $\varepsilon$ and $\rho$ in our model, we find $\bar{s}>0.05$. 

To estimate $\bar{s}$, we must consider three types of characteristic behavior possible in the demand-driven regime: noise-driven, limit cycle and coherence resonance behavior. Noise-driven behavior prevails when feedback is weak. This situation is, in its limit, equivalent to that of the supply-driven regime in which sentiment behaves symmetrically with respect to the origin. Therefore, $\bar{s}\rightarrow0$. Thus, the noise-driven mode generates growth $y_0\rightarrow\varepsilon$, which is lower than $R$.

The growth in the two other modes is studied numerically in Section \ref{sec_results_asymp}. For completeness, we briefly note, first, limit cycles (periodic or stochastic) lead to $\bar{s}\rightarrow0$ and $y_0\rightarrow\varepsilon$ (as the economy tends to spend a half of its time in the region where $s>0$ and the other half where $s<0$) and, second, coherence resonance yields $\bar{s}>0.05$ and $y_0>R$, owing to the attractors' asymmetry caused by technological growth ($\varepsilon>0$) in the presence of economic feedback ($\gamma>0$). 

As a final remark, it follows from \eqref{eq_lt_demand_y0} that, asymptotically, $z\sim z_0t\sim(\rho k_{d0}+\varepsilon - y_0) t \sim  O(1)$. The system's motion is therefore bounded in $z$. Its motion is likewise bounded in $s$ and $h$, which vary between -1 and 1, as, at the boundaries, $\dot{s}$ and $\dot{h}$ are directed into the domain of motion as follows, respectively, from equations \eqref{eq_res_full_s_dot} and \eqref{eq_res_full_h_dot}. Thus, the system's phase trajectories are bounded in the $(s, h, z)$-space.

\end{document}